\documentclass[11pt,]{article}
\usepackage{amssymb,amsmath}
\usepackage[utf8]{inputenc}
\usepackage[margin=1in]{geometry}
\usepackage[disable]{endfloat}

\usepackage{hyperref}
\usepackage{longtable,booktabs}
\usepackage{graphicx}

\usepackage{color}

\makeatletter
\def\maxwidth{\ifdim\Gin@nat@width>\linewidth\linewidth\else\Gin@nat@width\fi}
\def\maxheight{\ifdim\Gin@nat@height>\textheight\textheight\else\Gin@nat@height\fi}
\makeatother
\setkeys{Gin}{width=\maxwidth,height=\maxheight,keepaspectratio}

\providecommand{\tightlist}{%
  \setlength{\itemsep}{0pt}\setlength{\parskip}{0pt}}
\let\rmarkdownfootnote\footnote%
\def\footnote{\protect\rmarkdownfootnote}

  \title{Bayesian probabilistic models for corporate context, with an application to internal audit activities}
  \author{Francesco Toraldo \thanks{Poste Italiane S.p.A., Rome, Italy}  \and Fabio S. Priuli \thanks{Bain \& Company Inc., Rome, Italy} }
  \date{\today}

\begin{document}
\maketitle
\begin{abstract}
In this paper we present a business case carried out in Poste Italiane, in the context of fair performance evaluations of human resources engaged in internal audit activities.
In addition to the development of a Bayesian network supporting the goal of the Internal Audit unit of Poste Italiane, the work has led to the development of a methodological approach to advanced analytics in corporate context, whose usefulness goes well beyond the specific use case described here.
We thus present the different stages of such analytical strategy, from feature selection, to model structure inference and model selection, as a general toolbox that allows a completely transparent and explainable process to support data-driven decisions in business environments.
\end{abstract}

\hypertarget{introduction}{%
\section{Introduction}\label{introduction}}

Data analytics and machine learning represent a resource with ever growing
presence in corporate context. Mainstream media interest seems to be
captured only when a new advancement is made in the field of computer vision
and image recognition, or when an intelligent algorithm defeats humans in
some board games, or when a scandal uncovers misuses of private data
collected by big IT companies. In fact, the potential contribution of data
science in moving corporate decisional processes towards a data-driven path
has made them an invaluable tool in a much wider variety of contexts: from
capacity planning to fraud management, from real time production monitoring
to autonomous driving and controlling systems.

In this article, we propose an (almost) \emph{end-to-end chain of data-driven solutions},
designed to support the various phases of typical corporate decision making
processes, through an extensive usage of Bayesian probabilistic tools. Such
methodological framework allows to treat a wide range of business processes
with almost no need of \emph{ad hoc} adjustment, while ensuring coherent
evaluations of uncertainty and enabling several kinds of \emph{what if} analysis.
Our proposed framework is the byproduct of a series of projects developed by
the Internal Audit department of Poste Italiane in 2015-2019, in strict
collaboration with data scientists from an external advisory company Pangea
Formazione, now part of Bain \& Company. Each project was built on top of the earlier ones, expanding the
scope of the probabilistic forecast to cover further analysis, which allowed,
in turn, more informed decisions. In particular, the activities developed in
the last year led to the completion of the framework we present here: a few
software solutions which could support effectively the whole decision making
process, from the selection of relevant variables, to the formulation of a
suitable statistical model of the process of interest, to the actual
inference of parameters through some of the most advanced statistical
techniques available, and to the prediction of variables of interest.

In \protect\hyperlink{secEX}{Section \ref{secEX}} we will give a detailed outline of a motivating
application, taken from the business case that involved experts from the
Internal Audit unit at Poste Italiane: a tool to monitor performance
evaluations of auditors based on past evaluations of ``similar'' auditors, so
to highlight possible deviations and personal biases in such evaluations
and, in case, to ensure more fair judgements. However, it shall be clear
that \emph{the same basic idea could be applied well beyond this specific example}:
Poste Italiane itself has
applied this approach in multiple internal audit activities and to monitor
risk impacts of changes in procedures and controls; and Pangea Formazione
has implemented analogous tools in a variety of corporate contexts, from
capacity planning to marketing strategies, from predictive maintenance to
retention and anti-churn activities.

In general, in presence of complex corporate procedures and activities,
Bayesian tools allow to synthetize all available information and to exploit
them in a consistent framework, while retaining the capability to inspect in
depth the reasons behind each evaluation (through the inspection of specific
marginal distributions, given some evidences and observations). This offers
a huge advantage, in large business activities, compared to other machine
learning solutions like \emph{boosting trees} or \emph{deep neural network} which act
more like ``black boxes'', because it gives decision makers a mean to \emph{explain}
the rationale behind their decisions.
At the same time, the procedure does not completely automatize the chain of
operations: the quantitative results provided at each step shall be (and
typically are) reviewed by human experts, to monitor the possible presence of
biases in the collected samples of data, or to integrate effects that
experts know are present but are for some reason hard to observe in the
data available.
This integration of human and algorithmic capabilities tries to take the best
of both worlds (AI as \textbf{augmented intelligence} rather than
\textbf{artificial intelligence}) and it allows to validate the proposals of
each algorithm through the experience of people who work daily on the
processes of interest.

In the following sections, we will give more details about the four different phases that constitute our approach and that provide support to different stages of corporate decision making processes:

\begin{itemize}
\item
  Phase 1 - \emph{variable selection} based on the concepts of mutual information and conditional mutual information;
\item
  Phase 2 - \emph{structure inference} of Bayesian networks for the process of interest, combinining the variables selected in the earlier phase with experts' proposals;
\item
  Phase 3 - \emph{model comparison} of the more promising models through Bayes-Turing factor, and comparison with a baseline benchmark (typically a na\"{\i}ve Bayes classifier, if we are dealing with discrete classification problems);
\item
  Phase 4 - \emph{parameter inference and predictions} through \emph{Markov chain Monte Carlo} simulations (\emph{MCMC} in the following).
\end{itemize}

While each step is not original, the actual application of the whole
procedure in corporate contexts is far from common, to our knowledge, and no
commercial data analysis or statistical platform seems to implement it
organically. Thus, in the projects at Poste Italiane, we developed our own
tools in the form of scripts written in the statistical programming language
R~\cite{ref-Rlang}, by incorporating several of its statistical 
libraries~\cite{ref-ScutariBN, ref-BNCL, ref-SuzukiBN} and 
by exploiting for the MCMC simulations
the JAGS library, through its R interface~\cite{ref-PlummerRJAGS}. Such a solution
allows a flawless integration in every business workflow where it is allowed
to call external command line scripts as an intermediate stage, with
resulting structures and data saved to temporary files.

\hypertarget{secEX}{%
\section{An example of corporate application}\label{secEX}}

The Internal Audit group in Poste Italiane has been using Bayesian modelling
techniques to perform quantitative evaluation and scenario analysis since a
few years. They applied such techniques in very different contexts, both as
a tool to turn different sources of evaluations into more coherent
indicators and as a tool to forecast multiple KPIs for audit activities,
based on the characteristics of each post office and of each team
of auditors.

Bayesian techniques are invaluable in corporate processes, because they allow 
to exploit all available information, usually quite sparse, and to synthetize it 
consistently with experts' opinions that can be modelled, not without considerable
efforts, in forms of prior distribution of probability to be inserted into the model.

In this section, we present a concrete example in which our proposed \emph{end-to-end} 
procedure was applied at Poste Italiane. Namely, we built a Bayesian model which aimed to 
support performance evaluations of human resources engaged in internal audit activities, 
after having learnt from historical data what kind of evaluation `similar' auditors
had received.
The construction was performed through the steps outlined in \protect\hyperlink{introduction}{Section \ref{introduction}}: 
by selecting relevant variables through their mutual information, by 
inferring different Bayesian networks structures and comparing them through subject matter's 
opinions and the Bayes-Turing factor \emph{BF} (whose logarithm is also known as the ``weight of evidence''),
by performing predictions of the quantity of interest through MCMC simulations.

The plan of this section is, therefore, to start describing the context and
the relevant features that determine the quantities at play, then to move
to describe each step of the construction in more details. 
Since we are not adding new statistical developments, we defer to 
the \protect\hyperlink{appTech}{Appendix \ref{appTech}} the description of 
some of the statistical tools and techniques used throughout the process.

A few key aspects of the construction are the following:

\begin{itemize}
\tightlist
\item
  for all network structures, we make the same assumptions on the data-generating process by assigning a Categorical distribution to each feature, and assuming their parameters are \emph{a priori} uniformly distributed;
\item
  we compare structure learning algorithms which are built by optimizing different scores (BIC, likelihood or BDeu), by using both \emph{BF} \eqref{eq:BF} and the predictions on a separate test set (to measure latter predictive performances, we use \emph{root of mean square error} because we primarily aim to small errors);
\item
  in the application at hand, the model selection phase through \emph{BF} does
  not lead directly to the selection of a single candidate model: since we have
  only a few hundreds of data records available, in order to
  mitigate the risk of overfitting, we keep all models that
  perform better than a na\"{\i}ve Bayes classifier, and then we perform a 10-fold cross
  validation procedure, along the lines of~\cite{ref-BayesCV}, to make our final
  choice.
\end{itemize}

In view of the focus on categorical features, we make large use of information measures (like Shannon entropy and mutual information)
both to select variables to be included in the model, and to review different network structures, in search for links that could be safely removed
from the model. In presence of continuous features, or mixtures of both continuous and discrete variables, such quantities require to
assume a suitable model of the joint distribution of pairs of variables. When subject matter experts want to avoid such a \emph{modelling step},
it becomes important to exploit other available ``correlation'' measures or selection approaches: e.g. by using nonparametric sure independence 
screening~\cite{ref-FanLv08, ref-FanFS11} to decide which features can be discarded; by choosing suitable prior distributions to continuous parameters, 
so to penalize models with large parameters or too many variables, like with ridge or LASSO regularization~\cite{ref-BayesLASSO}; by measuring the impact of 
each independent variable on the target through partial dependency plots, or ICE plots, similarly to what could be done with black box models~\cite{ref-BBOXzh}.
We do not account for these alternative approaches in this paper.
 
\hypertarget{corporate-context-internal-audit-activities}{%
\subsection{Corporate context: internal audit activities}\label{corporate-context-internal-audit-activities}}

The motivation behind the construction of a model for performance
evaluations of human resources is to ensure more homogeneous judgements,
through identification of possible biases in the auditors' evaluations.
Namely, a predictive model that suggests an appropriate score to
different auditor profiles is expected to provide a solid \emph{anchor} to
compare further evaluations with, so to avoid the risk of making them too
subjective.
Evaluations given by the model would not be prescriptive, but they would help to highlight
deviations from the past judgements, making possible to analyse whether such deviations are really motivated
by changes in the priorities of the corporation, that might e.g.~decide at some point
to give priority to reward more innovation skills, or if it is affected by external factors that should not
have influence. Most often the reason for non-conforming evaluations is a
combination of different factors, and the model can be used to raise an alarm whenever there is a difformity that shall be better understood.

In order to decide what kind of profile each auditor would belong to, we use a set of
nine predictors, extracted from the \emph{Leadership model} used in Poste Italiane
and linked to the conceptual category of ``abilities''. Such predictors represent measures
for the evaluation of all workers in Poste Italiane, not only auditors. They are in detail:

\begin{itemize}
\tightlist
\item
  \textbf{Value for the customer}: a measure of the centrality of the customers in the activities, based on the rapidity of the services offered and the attention to customer's needs;
\item
  \textbf{Innovation and simplification}: a measure of the capacity to exploit innovative approaches and to simplify procedures, while keeping a high level of quality;
\item
  \textbf{Vision}: a measure of the capacity to read the whole context in which daily activities are embedded, acknowledging interdependencies among processes and their complexity, while suggesting solutions that could help operativity;
\item
  \textbf{Propensity to change}: a measure of the capacity to adopt new operative models that are being progressively introduced in the company, and to support transformations through direct involvement and collaboration;
\item
  \textbf{Decision-making ability}: a measure of the skills related to properly assessing risks and future scenarios and to effectively managing uncertainties in a consistent and appropriate way;
\item
  \textbf{Inclusive leadership}: a measure of the team leading ability and the level of involvement established with colleagues, collaborators and customers;
\item
  \textbf{Integrity}: a measure of the respect of ethical principles with coherent behaviors both in daily operativity and in the interactions with customers and colleagues;
\item
  \textbf{Collaboration and ability to influence}: a measure of the capacity to cooperate with other business units in order to improve and speed up decision making processes;
\item
  \textbf{Resource Engagement}: a measure of the level of engagement and proactivity of the resource in her daily job.
\end{itemize}

In the leadership model, these quantities are evaluated as an integer number in
the range 1 to 10. However, when used in the evaluation of auditors, the
features are discretized into quintiles in order to facilitate their
handling. Based on historical data, Internal Audit experts have chosen 5
thresholds for each features, so that data records are split into 5 equally
numerous buckets. Then, present and future auditors are assigned an integer
score from 1 to 5 depending on the interval their numerical score belongs
to: 1 if they are in the top 20\% of the evaluations, and then decreasing
towards 5 if they are in the bottom 20\%.

The target variable for the model is then the ``global evaluation'' for a new auditor (named
\textbf{Final evaluation} in our dataset), based on its characteristics and attitudes, as an integer in a scale from 1 to 10, with 1 being the best value and 10 being the worst one.

For the analysis, we had historical data related to \(234\) auditors: the small number is due to the need for only using \emph{fresh} data in
the model, because of the moving objectives of the criteria used, which tend to be refined
as time goes. As a result, evaluations of skills from a few years ago are not really comparable with newer ones.
This restriction has an unfortunate side effect, in this specific application,
because the target variable, even though it is supposed to span from 1 to 10, is only present with values in 1 to 8
because no records with the worst scores are included. This
means that the model is not expected to have high precision for very negative cases, and
that MCMC simulations will often require more iterations to reach convergence for the
target Dirichlet distributions.

\hypertarget{variable-selection}{%
\subsection{Variable selection}\label{variable-selection}}

The first step consists of selecting the most promising predictors, based on mutual information scores (both unconditional and conditional) as defined in \eqref{eq:normalized}. A series of R scripts, applied to
our original dataset, allows to create different scoring tables like in Table \ref{tab:scoreTableGeneric}:

\begin{itemize}
\tightlist
\item
  a table with the values of \(\mathrm{MI}'(X,Y)\), for each pair of predictors \((X,Y)\), sorted by decreasing scores (see Table \ref{tab:scoreTableGeneric}-top);
\item
  a table with the values of \(\mathrm{CMI}'(X,Y|Z)\), for each triple of predictors \((X,Y,Z)\), sorted by decreasing scores (see Table \ref{tab:scoreTableGeneric}-middle);
\item
  a table with the comparison among the scores \(\mathrm{MI}'(X,Y)\) and \(\mathrm{CMI}'(X,Y|Z)\), so to highlight situations where pairs of variables which would usually be dropped, could instead be highly informative, when their distribution is considered as conditional on a
  third variable (see Table \ref{tab:scoreTableGeneric}-bottom).
\end{itemize}

\begin{table}
\caption{\label{tab:scoreTableGeneric}Sample of rows from the tables containing mutual information scores \eqref{eq:normalized} which are relevant for our feature selection process: the first table contains the $\mathrm{MI}'$ score for each pair of features, in decreasing order; the second table lists the $\mathrm{CMI}'$ score for each triple of features, in decreasing order; finally, the third table presents the variation (delta and perc) between $\mathrm{MI}'$ and $\mathrm{CMI}'$ when one assumes conditioning on a third variable.}

 \centering

\begin{tabular}{c}

\\~\\

\begin{tabular}{ll}
 \toprule
pair & \(\mathrm{MI'}\)\\
 \midrule
\(\mathrm{MI'}\)(PROP\_TO\_CHANGE, COLLAB\_INFL) & 0.281\\
\(\mathrm{MI'}\)(LEAD\_INCL, COLLAB\_INFL) & 0.271\\
\(\mathrm{MI'}\)(VISION, PROP\_TO\_CHANGE) & 0.27\\
\(\mathrm{MI'}\)(PROP\_TO\_CHANGE, LEAD\_INCL) & 0.25\\
\(\mathrm{MI'}\)(VISION, COLLAB\_INFL) & 0.249\\
\(\mathrm{MI'}\)(PROP\_TO\_CHANGE, DEC\_MAKING) & 0.247\\
 ... & ...\\
 \bottomrule
 \end{tabular}

\\~\\
\\~\\

\begin{tabular}{ll}
 \toprule
triplet & \(\mathrm{CMI'}\)\\
 \midrule
\(\mathrm{CMI'}\)(PROP\_TO\_CHANGE, COLLAB\_INFL | RES\_ENG) & 0.3\\
\(\mathrm{CMI'}\)(VISION, PROP\_TO\_CHANGE | RES\_ENG) & 0.288\\
\(\mathrm{CMI'}\)(LEAD\_INCL, COLLAB\_INFL | RES\_ENG) & 0.274\\
\(\mathrm{CMI'}\)(PROP\_TO\_CHANGE, DEC\_MAKING | RES\_ENG) & 0.271\\
\(\mathrm{CMI'}\)(PROP\_TO\_CHANGE, LEAD\_INCL | RES\_ENG) & 0.266\\
\(\mathrm{CMI'}\)(INNOV\_SIMPL, LEAD\_INCL | RES\_ENG) & 0.261\\
 ... & ...\\
 \bottomrule
 \end{tabular}

\\~\\
\\~\\

\begin{tabular}{lllllll}
 \toprule
x & y & z & \(\mathrm{CMI'}\) & \(\mathrm{MI'}\) & delta & perc\\
 \midrule
 INTEGRITY & RES\_ENG & VAL\_FOR\_CLI & 0.048 & 0.026 & 0.021 & 80.3 \%\\
 VAL\_FOR\_CLI & RES\_ENG & INTEGRITY & 0.044 & 0.024 & 0.019 & 79.1 \%\\
 INNOV\_SIMPL & RES\_ENG & INTEGRITY & 0.039 & 0.023 & 0.016 & 68.3 \%\\
 INTEGRITY & RES\_ENG & DEC\_MAKING & 0.044 & 0.026 & 0.018 & 66.4 \%\\
 INTEGRITY & RES\_ENG & INNOV\_SIMPL & 0.044 & 0.026 & 0.017 & 65.1 \%\\
 DEC\_MAKING & RES\_ENG & INTEGRITY & 0.043 & 0.027 & 0.016 & 59.4 \%\\
 ... & ... & ... & ... & ... & ... & ...\\
 \bottomrule
 \end{tabular}

\end{tabular}

\end{table}

\begin{figure}

{\centering \includegraphics[width=0.45\linewidth]{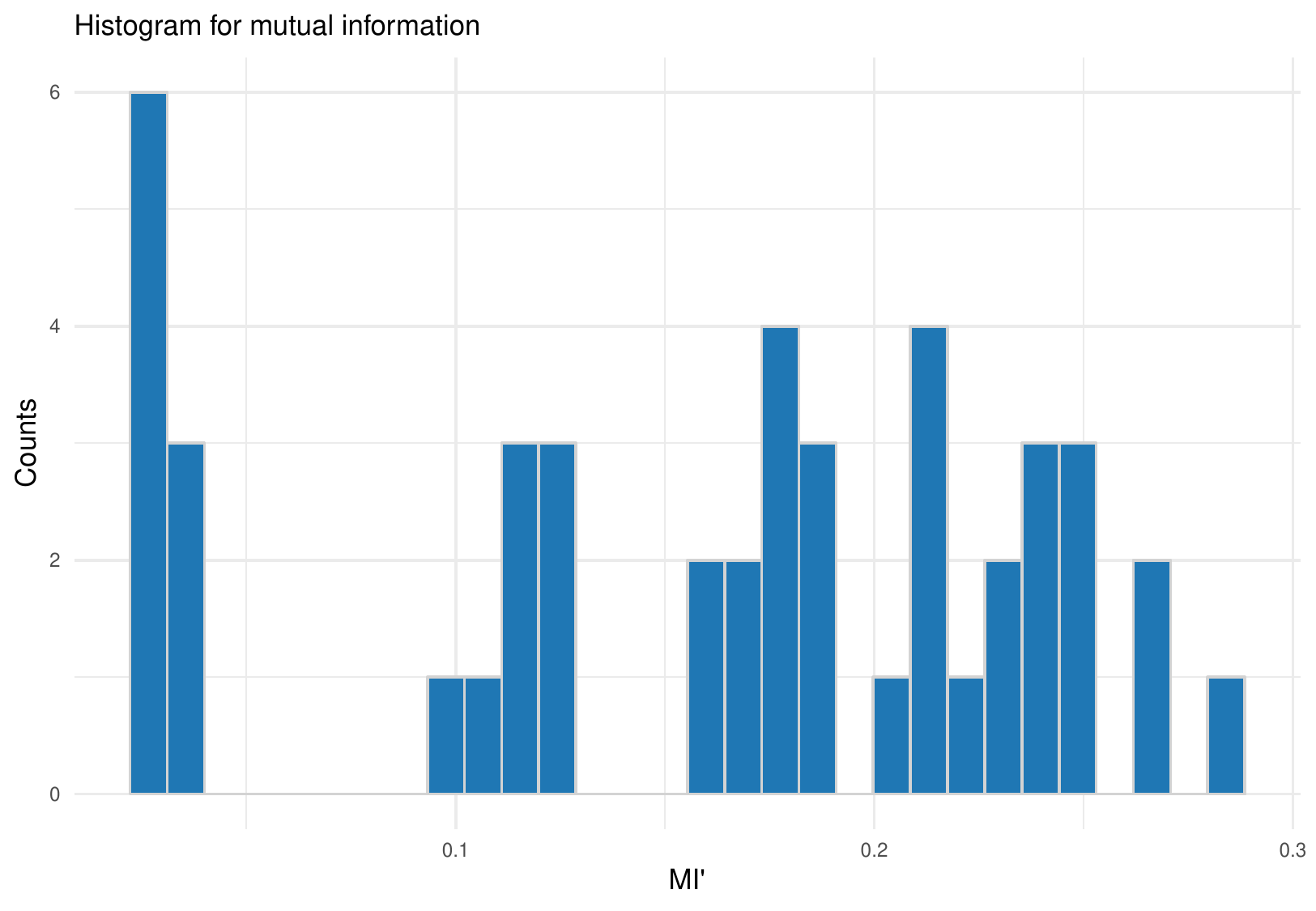} \qquad\includegraphics[width=0.45\linewidth]{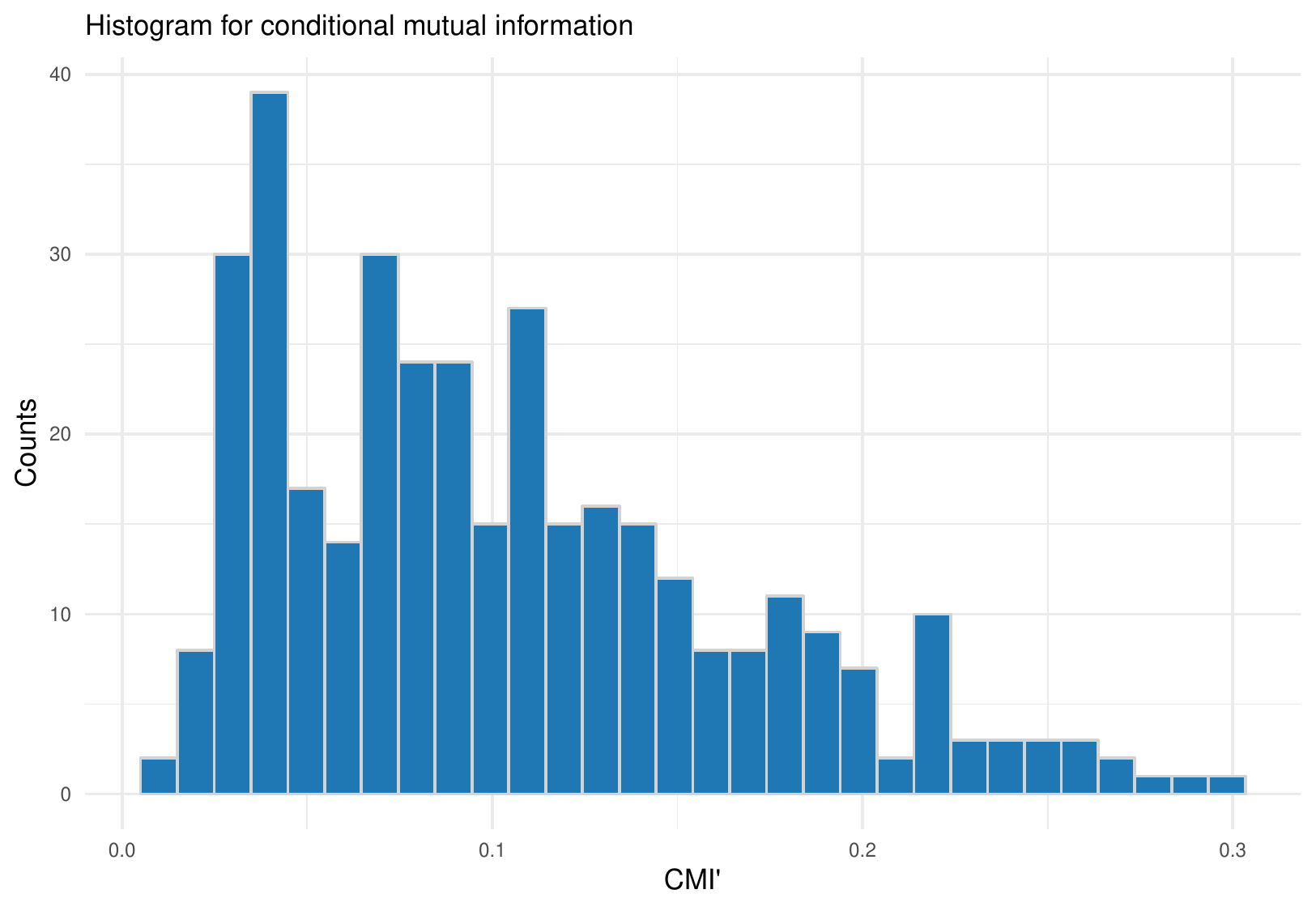} 

}

\caption{Distributions of normalized mutual information scores \eqref{eq:normalized}, unconditional (left) and conditional (right), can help the choice of an acceptance threshold in the feature selection process.}\label{fig:histogramsMI}
\end{figure}

They also provide histograms of the distributions of \(\mathrm{MI}'\) and \(\mathrm{CMI}'\) (cfr. Figure \ref{fig:histogramsMI}),
so to effectively support the understanding of how each score is positioned in the overall distribution and the possible choice of an acceptance threshold to drive feature elimination.

Recalling that at this stage we are mostly interested into information scores
from \eqref{eq:normalized}, so to select which features to include in our
predictive model, we focus our attention to the bottom of the \(\mathrm{MI}'\)
table, where less informative variables are located, and we analyse their
contribution.
In this specific case, the variable which have smallest mutual information with the others is
\textbf{Resource Engagement}, which is responsible for the tail of the histogram of \(\mathrm{MI}'\) in Figure \ref{fig:histogramsMI}-left. Before deciding whether to drop it, though, we want to check also the possible increase observed when the distribution of \textbf{Resource Engagement} is conditional on other variables, i.e.~its \(\mathrm{CMI}'\) score.
This is indeed the case, as one can see in Table \ref{tab:scoreTableGeneric}-bottom: \(\mathrm{CMI}'(X, \mbox{Resource Engagement}|Z)\) do increase when compared to \(\mathrm{MI}'(X, \mbox{Resource Engagement})\), for many choices of variables \(X\) and \(Z\). However, the values of \(\mathrm{CMI}'\) remain too small compared to the ones of the other variables, available on Table \ref{tab:scoreTableGeneric}-middle, and thus do not seem to justify the inclusion of the variable \textbf{Resource Engagement} in the model.
Similarly, focusing the \(\mathrm{MI}'-\mathrm{CMI}'\) comparison to the case when
\textbf{Resource Engagement} is the conditioning variable, as shown in Table \ref{tab:deltaTable}, we get an
increase in \(\mathrm{CMI}'\), but the variation compared to the corresponding
\(\mathrm{MI}'\) seems too small to justify the inclusion of the variable.

\begin{table}

\caption{\label{tab:deltaTable}Before discarding a feature (e.g.~\textbf{Resource Engagement}), one shall check the conditional mutual information scores \eqref{eq:normalized}, to be sure that such variable does not play a relevant role as parent node of other features: this can be evaluated by examining the variations between $\mathrm{MI}'$ and $\mathrm{CMI}'$.}
\centering
 \begin{tabular}{lllllll}
 \toprule
x & y & z & \(\mathrm{CMI'}\) & \(\mathrm{MI'}\) & delta & perc\\
 \midrule
 VAL\_FOR\_CLI & INTEGRITY & RES\_ENG & 0.126 & 0.097 & 0.029 & 29.8 \%\\
 INNOV\_SIMPL & INTEGRITY & RES\_ENG & 0.129 & 0.104 & 0.025 & 23.8 \%\\
 DEC\_MAKING & INTEGRITY & RES\_ENG & 0.152 & 0.127 & 0.025 & 19.9 \%\\
 VAL\_FOR\_CLI & PROP\_TO\_CHANGE & RES\_ENG & 0.176 & 0.156 & 0.02 & 12.8 \%\\
 INTEGRITY & FINAL\_EVAL & RES\_ENG & 0.129 & 0.115 & 0.014 & 12.4 \%\\
 PROP\_TO\_CHANGE & INTEGRITY & RES\_ENG & 0.142 & 0.127 & 0.015 & 11.8 \%\\
 ... & ... & ... & ... & ... & ... & ...\\
 \bottomrule
 \end{tabular}
\end{table}

Therefore, domain experts concluded against the inclusion of \textbf{Resource Engagement} in view of model parsimony.

\hypertarget{structure-learning-and-model-comparison}{%
\subsection{Structure learning and model comparison}\label{structure-learning-and-model-comparison}}

\begin{figure}

{\centering \includegraphics[width=0.48\linewidth]{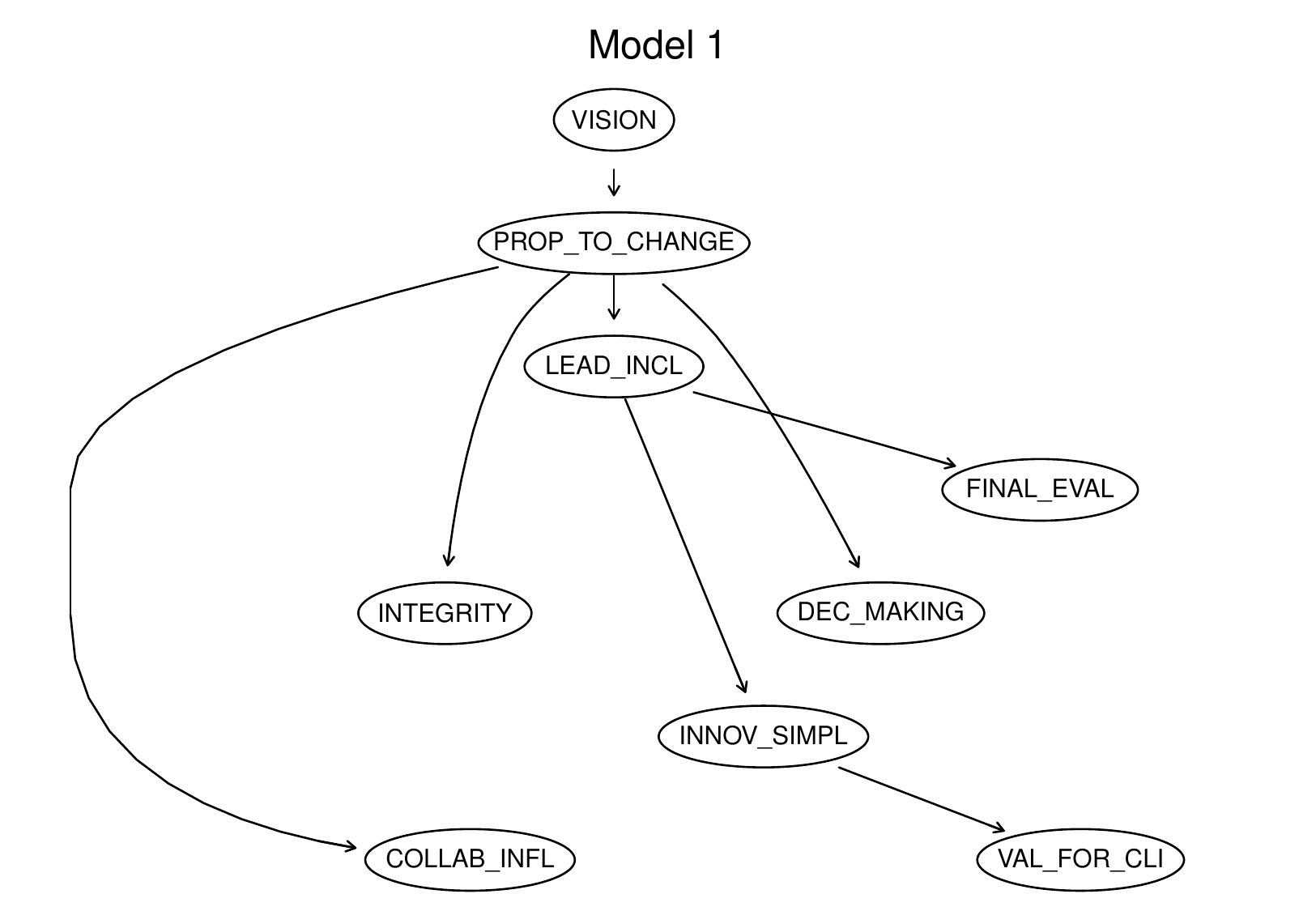} \includegraphics[width=0.48\linewidth]{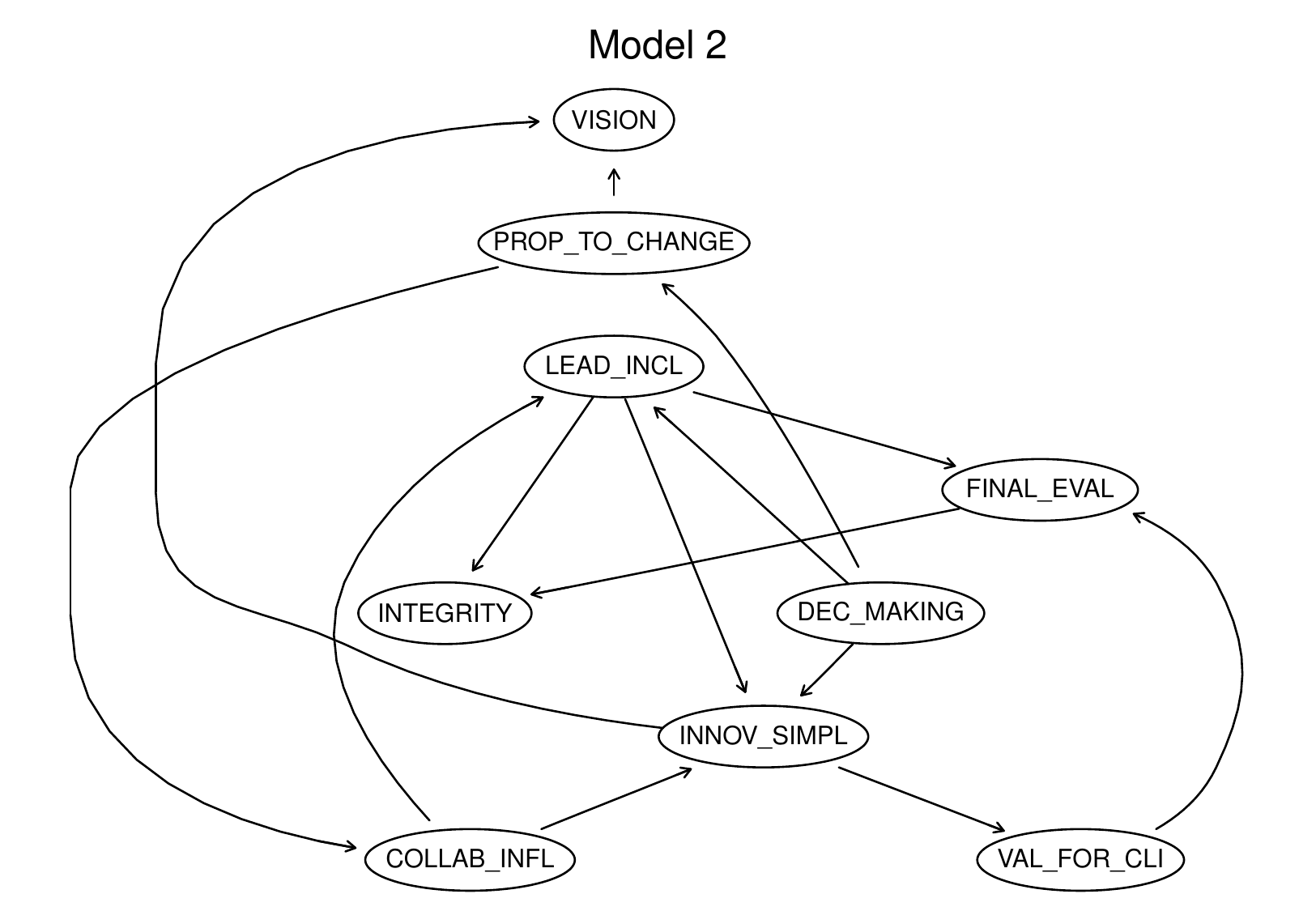} \includegraphics[width=0.48\linewidth]{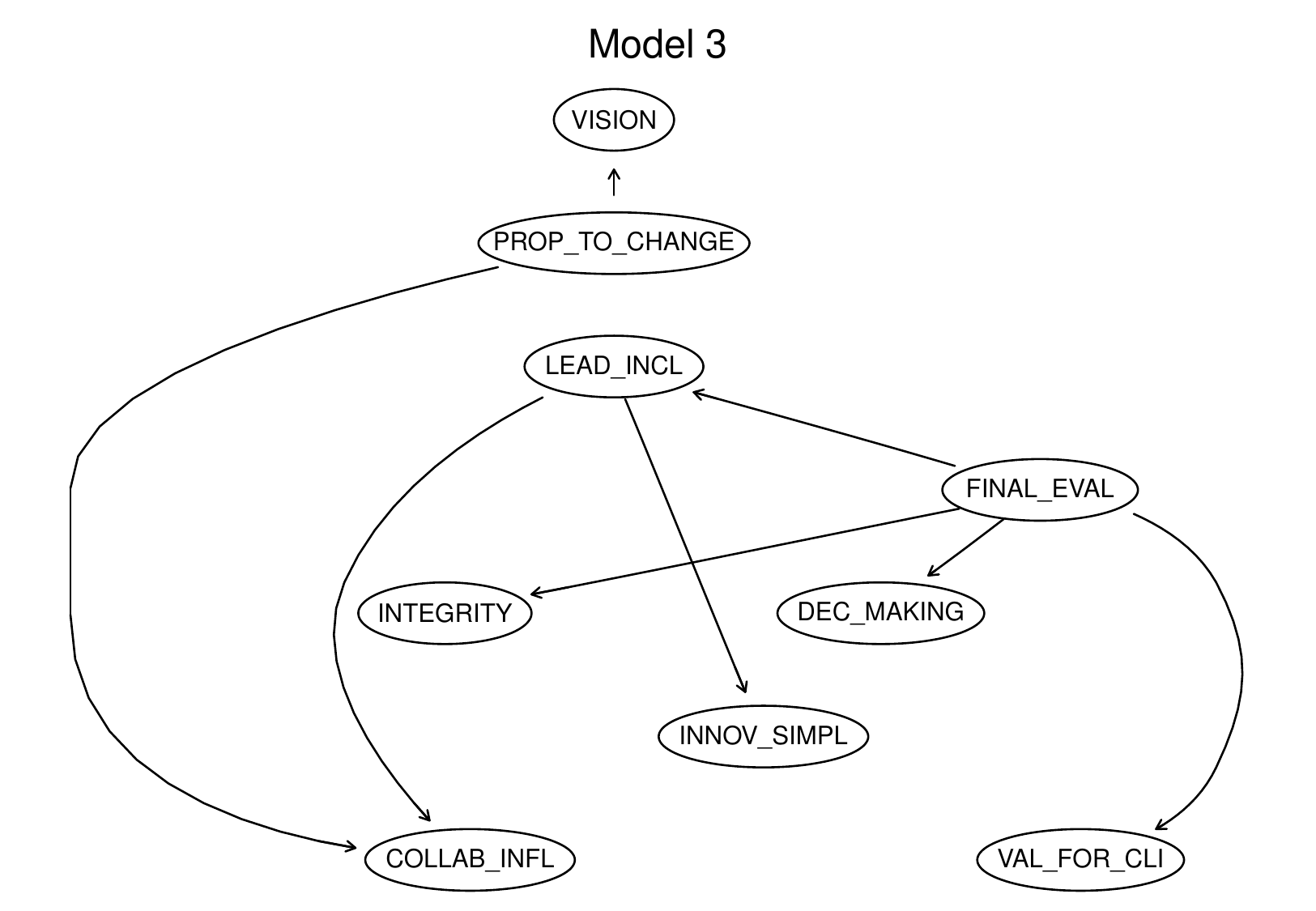} \includegraphics[width=0.48\linewidth]{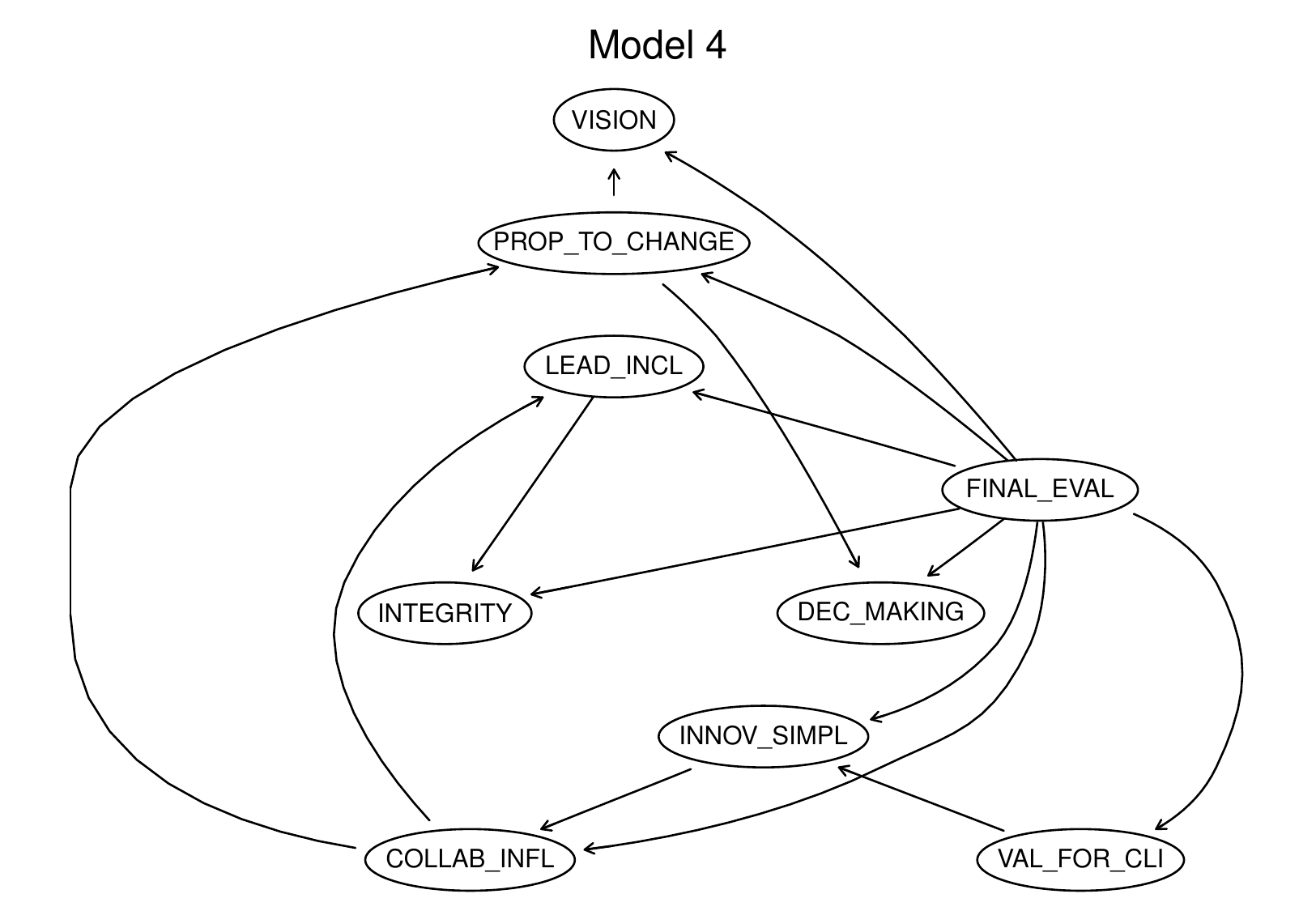} \includegraphics[width=0.48\linewidth]{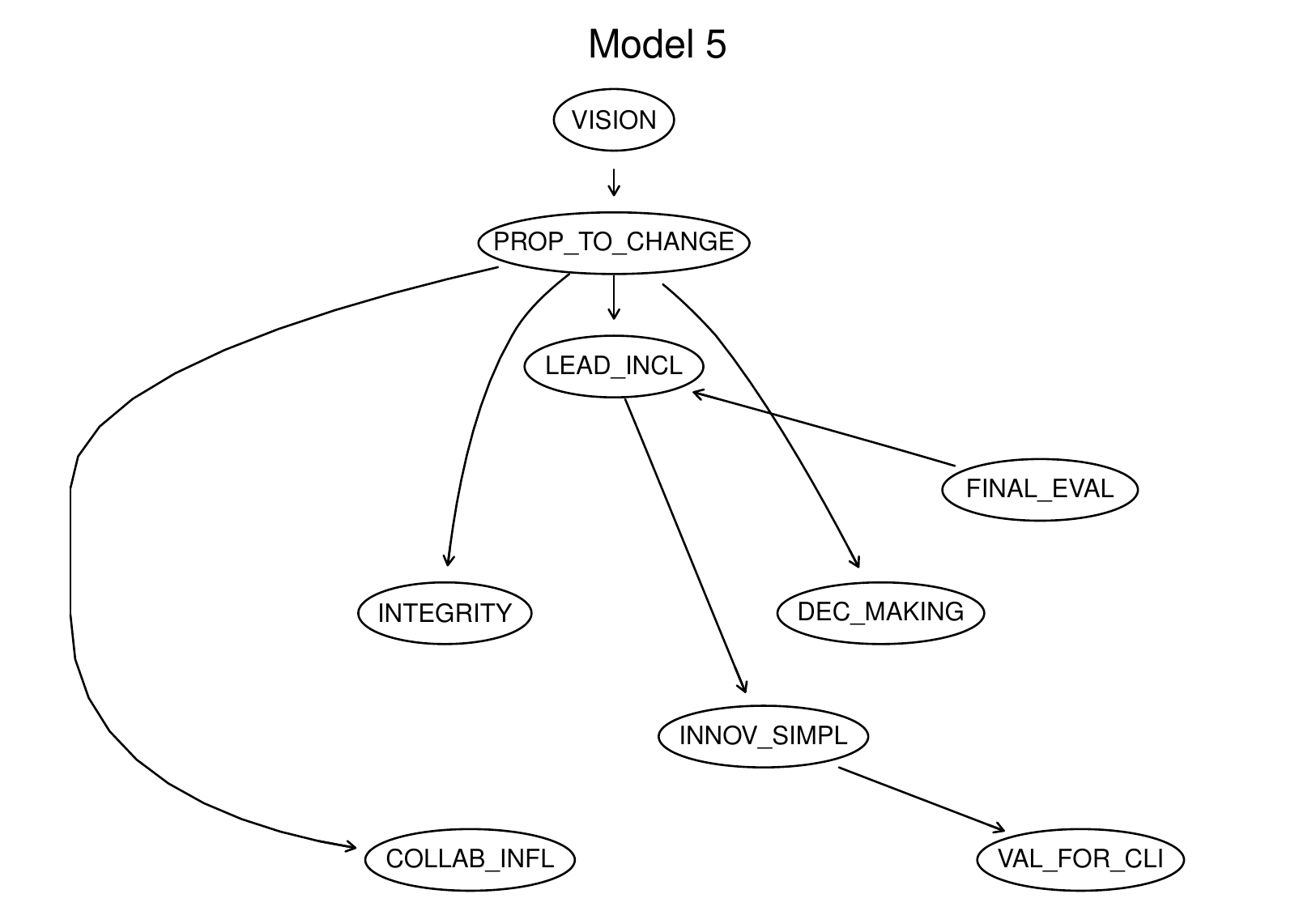} \includegraphics[width=0.48\linewidth]{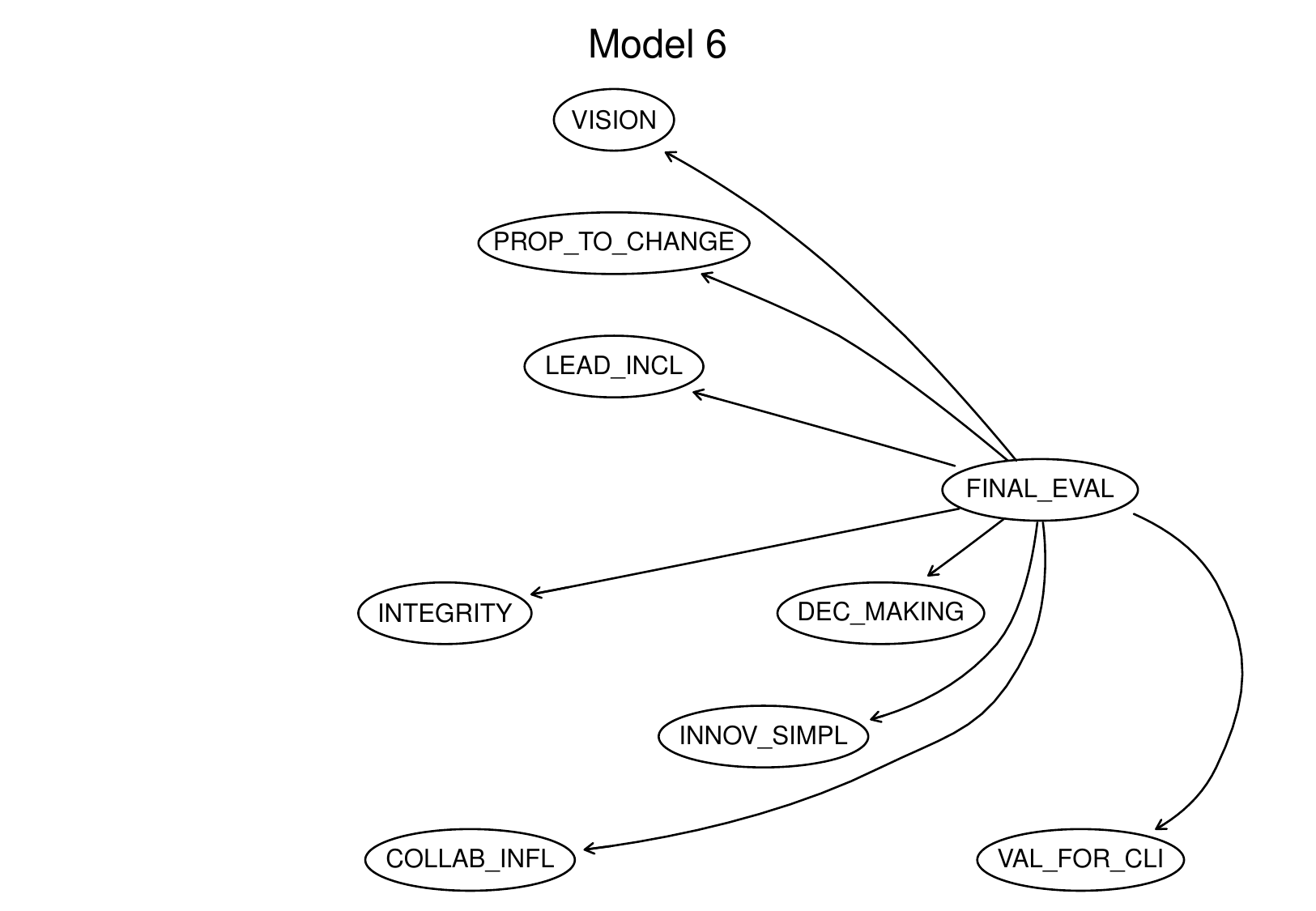} 

}

\caption{Different proposals for the connections between selected features. Here, we see \texttt{hc} (Model 1), \texttt{bnsl} (Model 2), \texttt{chowliu} (Model 3), \texttt{tan} (Model 4), \texttt{hc\_mod} (Model 5) and \texttt{naive} (Model 6). Notice how different algorithms sometimes switch direction in some connections (e.g.~\textbf{Vision} and \textbf{Propension to change}): experts can decide to impose logical constraints, so to prescribe one of the two directions over the other, but in this specific use case both connections were just accepted as valid, because experts considered both conditioning as reasonable.}\label{fig:compareBNstruct}
\end{figure}

Once the feature selection phase is completed, we move to the second stage
of our analysis, consisting of training several structure
algorithms over the available dataset.
The R scripts we use can automatically produce a series of reports for the expert to analyse.
Such reports typically include, for each available algorithm, the following components:

\begin{itemize}
\tightlist
\item
  proposal of network structure, as those in Figure \ref{fig:compareBNstruct};
\item
  a \emph{sensitivity analysis} of the target variable w.r.t. changes in every other predictors in the proposed model, as those in Table \ref{tab:compareSens}.
\end{itemize}

Here, by ``sensitivity analysis'' we mean a quantitative score given to the amount of information that each predictor \(V\) carries about the target variable \(T\).
An appropriate measure of such an impact is given by the variation in entropy \eqref{eq:entropy} due to the collected evidences, which we call \emph{sensitivity} \(\mathcal{S}(T, V)\) and define as:
\begin{equation}\label{eq:sensitivity}
\mathcal{S}(T, V) = H(T) - H(T|V).
\end{equation}
By measuring this quantity, experts can decide which connections to remove if the resulting network structure is still too rich of connections and thus too computationally expensive: the links between \(T\) and \(V\) with very small \(\mathcal{S}(T, V)\) can be removed.
Notice that in general the value \(\mathcal{S}\) is conditional on the graph
structure \(\mathcal{G}\) of the Bayesian network as well, even if we do not
make this dependence explicit through a notation like
\(\mathcal{S}_\mathcal{G}\).

In this application, we chose to focus our attention on a restricted selection of the available structure
learning algorithms in Scutari's \texttt{bnlearn} package~\cite{ref-ScutariBN}, Mihaljevic's \texttt{bnclassify}~\cite{ref-BNCL} and Suzuki's \texttt{BNSL}~\cite{ref-SuzukiBN}. Namely, we will compare the structures proposed from:

\begin{itemize}
\tightlist
\item
  ``Hill climbing'' greedy algorithm from \texttt{bnlearn}, denoted with \texttt{hc} in the following, which searches for a structure maximizing the Bayesian Information Criterion or \emph{BIC} (see~\cite{ref-ScutariBN} for details);
\item
  ``Tree augmented network'' algorithm from \texttt{bnclassify}, denoted with \texttt{tan} in the following, which maximizes likelihood as described in the paper by Friedman et al. (see~\cite{ref-FriedmanTAN} for details);
\item
  ``BNSL'' algorithm from the package with the same name, denoted with \texttt{bnsl} in the following, which contains Suzuki's implementation of Chow-Liu's algorithm and searches for structures which maximize a Bayesian Dirichlet (BD) score (see~\cite{ref-SuzukiBN} for details).
\end{itemize}

Most of these algorithms allow to enforce the presence of specific connections among variables or to ensure that some link should not be included. Some alternative tools also allow for the inclusion of logical constraints among variables or penalizations of larger cliques (see e.g.~HuginLite in~\cite{ref-HuginPaper}), if such input from expert is deemed worth the inclusion.
In the present case, we did not use any of these, and we preferred instead to include a couple of additional network structures which experts wanted to put under test:

\begin{itemize}
\tightlist
\item
  a modified version of the \texttt{hc} structure, where we invert the direction of the arrow connecting the variables \textbf{Final evaluation} and \textbf{Inclusive leadership}, denoted with \texttt{hc\_mod} in the following;
\item
  the undirected network structure proposed by ``Chow-Liu algorithm'' in \texttt{bnlearn}, with directions for the links added by the experts of
  the process, based on their experience and on some hints from the \(\mathrm{MI}'\) and \(\mathrm{CMI}'\) scores previously computed, and denoted with \texttt{chowliu} in the following.
\end{itemize}

Notice that, by changing some of the configuration options (= hyperparameters) in the algorithms above, e.g. the network score in \texttt{hc} or \texttt{tan}, 
we might obtain slightly different network structures, that shall be validated by domain experts, and possibly added to the list of models to be evaluated 
in the following. In the specific application considered here, though, no appreciable differences resulted by different choices of the hyperparameters.

In Figure \ref{fig:compareBNstruct} the five resulting network structures
are shown side-by-side for comparison, together with the structure of the
na\"{\i}ve Bayes classifier.
In both such additional structures the target variable,
\textbf{Final evaluation}, appears only as parent node for some of the remaining
variables. This is in line with the idea of using
measurable characteristics to infer the features that cannot be directly measured, like
the target one here. Such a choice also helps to keep under control the size of the conditional probability tables (CPT) of
the target variable, which is important in this specific example because of the small
size of the available dataset: as the number of parent nodes of a variable increases, indeed, so does
the number of parameter in its CPT, and therefore we risk to have not enough data to learn
their correct distributions.

\begin{table}
\caption{\label{tab:BFscore1}Sample of rows from the table containing the (logarithm of) BF scores \eqref{eq:BF}, for each pair of models (left) or for each model against na\"{\i}ve structure (right). The larger is BF(A,B), the most probable is model A compared to model B, given the available dataset.}

\centering
\begin{tabular}[t]{lll}
\toprule
model 1 & model 2 & log(BF)\\
\midrule
\texttt{hc} & \texttt{hc\_mod} & 105.977\\
\texttt{bnsl} & \texttt{hc\_mod} & 100.54\\
\texttt{hc} & \texttt{tan} & 97.428\\
 ... & ... & ...\\
 \bottomrule
 \end{tabular}
\centering \qquad\qquad \centering
\begin{tabular}[t]{lll}
\toprule
model 1 & model 2 & log(BF)\\
\midrule
\texttt{hc} & \texttt{naive} & 96.529\\
\texttt{bnsl} & \texttt{naive} & 91.092\\
\texttt{naive} & \texttt{hc\_mod} & 9.448\\
 ... & ... & ...\\
\bottomrule
\end{tabular}
\end{table}

Once we have a series of DAG proposals, our scripts compute Bayes-Turing factors \eqref{eq:BF} for each pair of models, and between the learned structures and the na\"{\i}ve one, as in Table \ref{tab:BFscore1}.
BFs can then be turned into an ordered hierarchy, as shown
in Table \ref{tab:orderedBFscore}, so to conclude that \texttt{hc} and \texttt{bnsl} are
definitely more capable to describe the dataset, compared to the other
algorithms. However, in this specific example, we had a very small dataset
and we were worried that the observed BFs might be affected by some
overfitting phenomena: recall that, due to its definition, the Bayes-Turing
factor selects the best DAG structure by itself, only when we can assume a
uniform probability
distribution among DAG structures. When domain experts find difficult to
discriminate a priori among different structures, we shall be careful in relying on the dataset alone for selection.
Here, we decided to exploit BF only to drop the model \texttt{hc\_mod} from the
analysis, since it scored definitely less than a na\"{\i}ve structure, and to postpone the final selection after a comparison of the predictive performances of
the remaining four models through a cross-validation procedure, which will be the focus of next section.

\begin{table}

\caption{\label{tab:orderedBFscore}From Table \ref{tab:BFscore1}, we can create an ordered rank among considered network structures: each model in the second column appears in the first column of next row, forming a chain of models in decreasing order of likelihood. Here, given the data, \texttt{hc} is more likely than \texttt{bnsl}, which is more likely than \texttt{chowliu}, etc.}
\centering
\begin{tabular}[t]{llr}
\toprule
model 1 & model 2 & log(BF)\\
\midrule
\texttt{hc} & \texttt{bnsl} & 5.437\\
\texttt{bnsl} & \texttt{chowliu} & 86.692\\
\texttt{chowliu} & \texttt{naive} & 4.401\\
\texttt{naive} & \texttt{tan} & 0.899\\
\texttt{tan} & \texttt{hc\_mod} & 8.549\\
\bottomrule
\end{tabular}
\end{table}

Before moving to the next phase, however, we want to stress that even at this stage it
is possible to perform an analysis of interest. Indeed, we can help experts in verifying their beliefs on the importance of different features for
the task at hand, against the actual sensitivity analysis offered by each network structure.
According to experts, the \textbf{Final evaluation} of auditors' job was mostly influenced by \textbf{Innovation and simplification},
\textbf{Integrity} and \textbf{Collaboration and ability to influence}. However, most of the structures
learned from the datasets suggest that, in fact, \textbf{Integrity} and \textbf{Collaboration and ability to influence} do not have a strong impact on the value of \textbf{Final evaluation}, while \textbf{Inclusive leadership} does.
In Table \ref{tab:compareSens} we show, side by
side, the sensitivity score \({\cal S}(T,V)\), as defined in \eqref{eq:sensitivity},
for our target variable \(T\) w.r.t. most relevant features \(V\), based on the
\texttt{hc} and \texttt{chowliu} network structures. As you can see, different DAG
structures change the impact of different predictors, but some conclusions
are shared across the DAGs.

\begin{table}
\caption{\label{tab:compareSens}By comparing the sensitivity scores $\mathcal{S}(T,V)$ from \eqref{eq:sensitivity}, relative to different network structures, experts were able to disprove their belief that $V = \mbox{\textbf{Integrity}}$ was very important in determining the target $T = \mbox{\textbf{Final evaluation}}$: in all structures, such variable had a limited impact on $T$.}

\centering \begin{tabular}{cc}

~\\

\begin{tabular}{lr}
\toprule
\multicolumn{2}{c}{\texttt{hc}}\\
\midrule
variable & \begin{tabular}[x]{@{}c@{}}sens. score\\ $\mathcal{S}(T,\cdot)$\end{tabular}\\
\midrule
LEAD\_INCL & 0.581\\
PROP\_TO\_CHANGE & 0.296\\
INNOV\_SIMPL & 0.283\\
COLLAB\_INFL & 0.190\\
VISION & 0.183\\
DEC\_MAKING & 0.170\\
VAL\_FOR\_CLI & 0.142\\
INTEGRITY & 0.078\\
\bottomrule
\end{tabular}

&

\begin{tabular}{lr}
\toprule
\multicolumn{2}{c}{\texttt{chowliu}}\\
\midrule
variable & \begin{tabular}[x]{@{}c@{}}sens. score\\ $\mathcal{S}(T,\cdot)$\end{tabular}\\
\midrule
LEAD\_INCL & 0.569\\
DEC\_MAKING & 0.491\\
VAL\_FOR\_CLI & 0.472\\
INNOV\_SIMPL & 0.271\\
INTEGRITY & 0.271\\
COLLAB\_INFL & 0.112\\
\bottomrule
\end{tabular}

\end{tabular}
\end{table}

Of course, this could be due to recent changes in the way such
characteristics have been evaluated in the leadership model (do not forget that
each feature in our dataset is not the direct measure of the skill of the
auditor, but it represents the quintile which each auditor belonged to,
based on its actual evaluation), or it could be due to some bias in the
limited dataset. In any case, it is an information that experts were highly
interested in, because one of their duties is to monitor in which ways
evaluations change in time, so to spot any unwanted distorsions that could
arise.

Notice that the usage of the sensitivity score~\eqref{eq:sensitivity} as a base for the considerations
above is again motivated by the presence of categorical variables only in our model. When dealing with continuous 
variables and parameters, or with a mixture of continuous and discrete ones, information theoretical quantities might be 
harder to compute than in the example we considered here. Thus, when this is the case, one can use some alternative metric
to compare the importance of variables in determining the predicted variables: for instance, by approximating 
the causal impact through partial dependency plots (assuming the model respects the underlying hypotheses), as in~\cite{ref-BBOXzh},
or by investigating contributions through techniques like \emph{LIME}~\cite{ref-LIME} or \emph{SHAP values}~\cite{ref-SHAP}.

\hypertarget{inference-of-parameters}{%
\subsection{Inference of parameters}\label{inference-of-parameters}}

\begin{figure}

{\centering \includegraphics[width=8.42in]{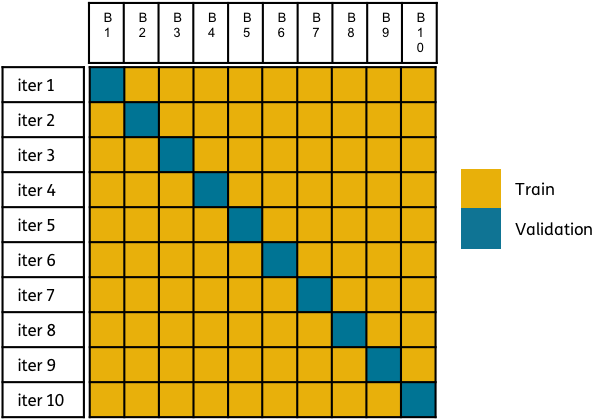} 

}

\caption{Cross validation scheme with 10 folds: each iteration of the cross validation procedure uses a different block as validation set and the remaining records as training set.}\label{fig:cv10}
\end{figure}

In the final step of our analysis, we want to choose the best model among
the four we had previously selected. In order to compare them all and to decide which one better suits our needs, we set up a
10-fold cross validation procedure to measure average performances of each model on 10
different training/validation tests, as depicted in Figure \ref{fig:cv10}.

In order to single out the best model and measure its predictive capabilities, we thus
divide our dataset in two different chunks of row indices: a training set comprising 85\%
of the data and a test set comprising the remaining 15\% of the records. The latter will
be only used to evaluate
the best model, once chosen, while the former will be extensively used in the cross
validation procedure.
Next, among the row indices of the training set, we select at random ten sets of \(23\)
indices, i.e.~10\% of the rows of the original dataset, for validation purposes: for each
run, the records corresponding to such indices will form the validation sets, while the
remaining data will be used as the actual training set for that run.

Hence, for each choice \(i=1,\ldots, 10\), we perform training of the four models by using
the \(i\)-th training set and we measure the performance on the \(i\)-th validation set.
Here, both training and predictions are performed through MCMC simulation, following the
ideas presented in \protect\hyperlink{inference-of-parameters-through-mcmc-simulations}{Section \ref{inference-of-parameters-through-mcmc-simulations}}. Nodes are all assigned a Categorical distribution, with as many parameters as the dimension of the CPTs. These parameters are initially assumed to be distributed uniformly
on the simplex with correct dimension (i.e.~with \(\mathrm{Dir}({\bf 1})\) for a vector \({\bf 1}\) of suitable dimension) and get updated during training. For each record of the validation set, the model assigns
as predicted `label' for \textbf{Final evaluation} the state with larger probability.
At the end of the cross validation procedure, we take the average of the ten performance
measures, in order to synthetize the results in a single score. In Table \ref{tab:cvPerf} we consider as measures:

\begin{itemize}
\tightlist
\item
  \emph{accuracy}, meant as the proportion of predictions which are correct over the total number of predictions;
\item
  \emph{root of mean square error} (\emph{RMSE} in the following), i.e.~the mean of the Euclidean distance between the vector of predictions and the vector of actual values of the target variable.
\end{itemize}

Notice that, here, by \emph{accuracy} we mean the elements on the diagonal of
the \emph{confusion matrix} for the classification problem
(such a matrix has in position \((i,j)\) the number
of records predicted of class \(i\) by the model while being
truly of class \(j\)).

\begin{table}

\caption{\label{tab:cvPerf}Performances of the four models included in the cross-validation procedure. In the evaluation table, we display both Accuracy (rate of correct classifications) and RMSE, but experts decided that the latter indicator was more useful because it also gave a measure of how far from correct predictions the model could go.}
\centering
\begin{tabular}[t]{lrr}
\toprule
Algorithm & Accuracy & RMSE\\
\midrule
\texttt{chowliu} & 0.413 & 0.769\\
\texttt{hc} & 0.165 & 0.921\\
\texttt{bnsl} & 0.204 & 1.019\\
\texttt{tan} & 0.217 & 1.053\\
\bottomrule
\end{tabular}
\end{table}

In this use case, experts decided that RMSE was the
most appropriate indicator of the two: indeed, we not only want a model that
makes few mistakes, but among models with equal accuracy we prefer the one
that, when wrong, is anyway as close as possible to the real value.
As such mean square error is more desirable than accuracy alone, because it penalizes large mistakes.
According to the comparison of the average RMSE, the best choice among the four models
was Chow-Liu's algorithm,
when equipped with conditional dependencies prescribed by
experts, i.e.~the structure we called \texttt{chowliu} (see ``Model 3'' in Figure \ref{fig:compareBNstruct}).

Having finally chosen the model we want to use, we retrain it on the complete training
set (85\% of the original data) and evaluate its overall performances on the test set
(15\% of the original data). Results are shown in Table \ref{tab:CLperf}.
Notice that, even if the model gives a wrong answer in about 2 out of 3 cases, it rarely
deviates more than \(1\), meaning that the answer can be effectively used as a support to
detect erroneous evaluation of auditors' job: even when the model makes mistakes, it is
anyway very close to the correct answer.

\begin{table}

\caption{\label{tab:CLperf}Summary of the performances for the selected model (\texttt{chowliu}) on the test set: even if the classifications are correct only in 1 out of 3 cases, the average error is smaller than 1, and only in 3 cases the error in the prediction is larger than 1.}
\centering
\begin{tabular}[t]{rrrrr}
\toprule
Cases & Correct & Error $\geq$ 1 & Accuracy & RMSE\\
\midrule
35 & 12 & 3 & 0.3429 & 0.8783\\
\bottomrule
\end{tabular}
\end{table}

One advantage of our Bayesian approach is that, in all stages, MCMC simulations produce complete distributions
of the categorical target variable \textbf{Final evaluation}
(cfr. Table \ref{tab:predictDistr1}), not only its most probable value.
This gives additional ``debugging'' tools to experts, compared to typical
pointwise estimates, since it offers a mean to verify how confident the
model is about its predictions, or to compare the expected values with
the true ones: if in a specific case the experts think the model might have
made a mistake, they can inspect the corresponding distribution and react
accordingly, see e.g.~Table \ref{tab:predictDistr2}.

\begin{table}

\caption{\label{tab:predictDistr1}Posterior categorical distributions for \textbf{Final evaluation}, as computed by MCMC simulations, for a sample of records in the test set, together with the mean evaluation, the predicted value (i.e.~the mode of the posterior) and the true value.}
\centering
\begin{tabular}[t]{lllllllllllll}
\toprule
1 & 2 & 3 & 4 & 5 & 6 & 7 & 8 & 9 & 10 & mean & predicted & true\\
\midrule
0.03 & 0.01 & 0.15 & 0.99 & 23.94 & 70.8 & 3.82 & 0.1 & 0.07 & 0.1 & 5.7813 & 6 & 5\\
0.07 & 0.03 & 0.01 & 0.01 & 0.12 & 24.76 & 74.47 & 0.3 & 0.13 & 0.11 & 6.7532 & 7 & 6\\
0.01 & 0.01 & 0.68 & 51.48 & 45.53 & 2.24 & 0.01 & 0.02 & 0.01 & 0.01 & 4.495 & 4 & 4\\
0.01 & 0.01 & 0.79 & 51.29 & 45.69 & 2.13 & 0.02 & 0.04 & 0.01 & 0.01 & 4.4944 & 4 & 5\\
... & ... & ... & ... & ... & ... & ... & ... & ... & ... & ... & ... & ...\\
\bottomrule
\end{tabular}
\end{table}

\begin{table}

\caption{\label{tab:predictDistr2}We show a few cases where 
consultants were in doubt due to apparently conflicting values of the 
predictors: experts would like to know how much they can trust the result 
given by the model. Thus, they look to the corresponding parameters of the 
distribution and observe that the model is way more convinced about its 
prediction in the first case, since state $5$ 
has about 62\% of probability to be the actual global evaluation, than in the 
second case, where 95\% of probability is distributed across 
the values $4$ to $6$, and $5$ is only slightly more probable than $6$. Notice 
that, in this example, even the expected value of the distribution would not 
be of great help and only inspection of the posterior distributions allows 
to discriminate the two situations.}
\centering
\begin{tabular}[t]{rrrrrrrrrrrr}
\toprule
1 & 2 & 3 & 4 & 5 & 6 & 7 & 8 & 9 & 10 & mean & predicted\\
\midrule
0.03 & 0.03 & 1.02 & 8.95 & 61.86 & 27.19 & 0.69 & 0.10 & 0.07 & 0.06 & 5.1825 & 5\\
0.03 & 0.04 & 1.86 & 25.37 & 35.99 & 34.77 & 1.67 & 0.18 & 0.03 & 0.06 & 5.0974 & 5\\
\bottomrule
\end{tabular}
\end{table}

We conclude this section by remarking that, in general, we expect the size of the split between training, validation and test
sets to have a moderate impact on the performance of some of the models. In the current example, though,
we saw no strong dependence on such a split, as long as enough records were kept for the training procedure:
while minor variations in the specific values for RMSE were noticed, when varying the size of the test sets 
between 10\% and 20\% of the records, the ranking of the performances of different models 
stayed basically the same; when increasing the size of the test set to around 25\% of the records, namely to 58 data entries, 
performances started to degrade for most models, due to underfitting because only \~150 records (65\% of the available data)
were used for training and some of the dependencies were not properly learned.

\hypertarget{details-about-mcmc-simulation}{%
\subsection{Details about MCMC simulation}\label{details-about-mcmc-simulation}}

In the final stage of our analysis, we performed inference of the parameters
of the chosen model via MCMC simulations. With the R packages described so
far, there are two possible options to perform such a step, thanks to the
discrete nature of the model considered in
this specific case: on the one hand, we can use \texttt{bnlearn} functions \texttt{bn.fit}
to first train the Bayesian network to the dataset, and then \texttt{predict} to
perform predictions of \textbf{Final evaluation} on the test set; on the other
hand, we can write a JAGS model for the specific network structure
\texttt{chowliu}, ``Model 3'' in Figure \ref{fig:compareBNstruct}, and then to
perform MCMC simulations through its R interface \texttt{rjags}~\cite{ref-PlummerRJAGS} to
both learn distribution parameters in each node and predict
\textbf{Final evaluation} on the test set.

In both cases, we need to examine the behavior of the sampler (after the \emph{adaptation} and \emph{burn-in} phases of the algorithm), through \emph{trace plots} and \emph{density plots} of each parameter. Here, there
would be plenty of parameters to inspect, as listed in Table \ref{tab:nParams}, since each conditional
dependence added to a node increases of a factor \(5\) the number of parameters to
describe the Dirichlet distribution for that node (it is a factor \(10\) if the parent
node is the target variable).
We want to stress that, in general, the results of the inference might depend on the number of samples used in the simulation and in the number of Markov chains that
are sampled. These are additional hyperparameters that should be accounted for in a more complex analysis, especially with continuous parameters appearing into 
the Bayesian structure. In our example, with categorical features only, we have seen no instabilities by using the default numbers of samples for
adaptation and burn-in, and sampling more than three chains added no valuable insights to the inference.

\begin{figure}
{\centering \includegraphics[width=0.24\linewidth]{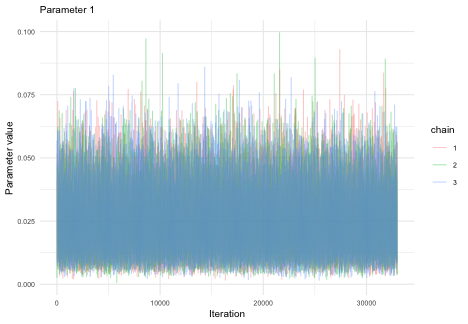} \includegraphics[width=0.24\linewidth]{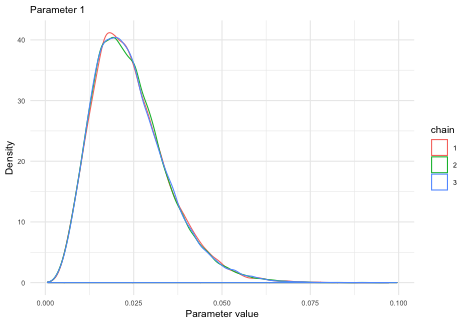} ~~\includegraphics[width=0.24\linewidth]{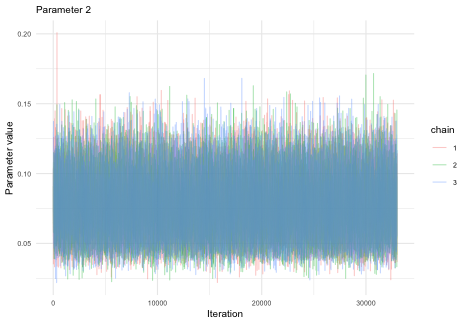} \includegraphics[width=0.24\linewidth]{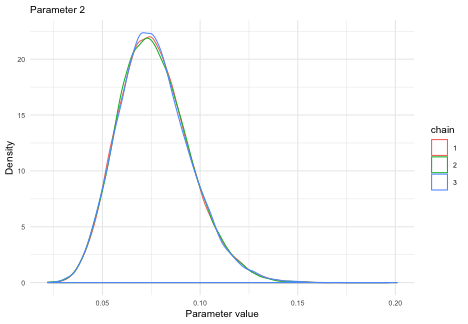} 

}

{\centering \includegraphics[width=0.24\linewidth]{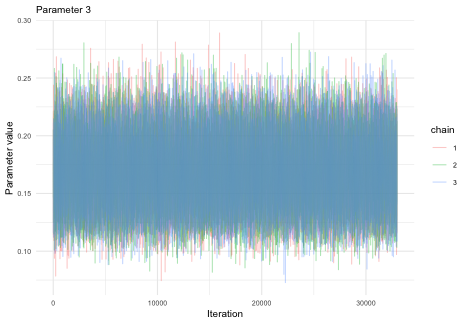} \includegraphics[width=0.24\linewidth]{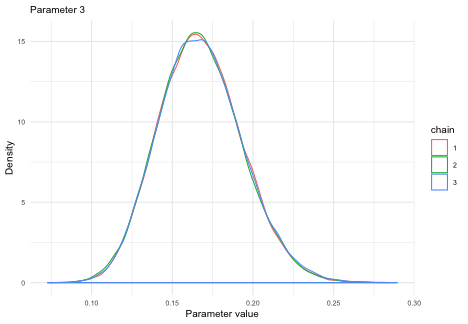} ~~\includegraphics[width=0.24\linewidth]{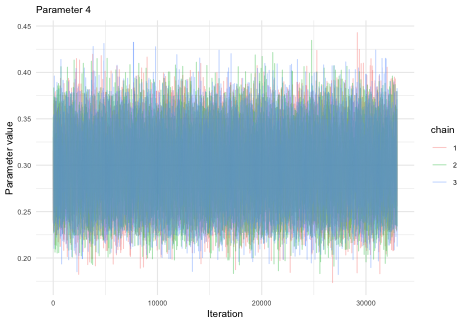} \includegraphics[width=0.24\linewidth]{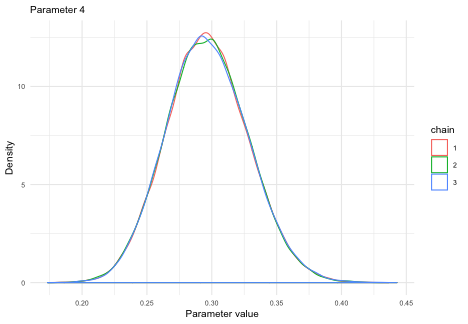} 
}

{\centering \includegraphics[width=0.24\linewidth]{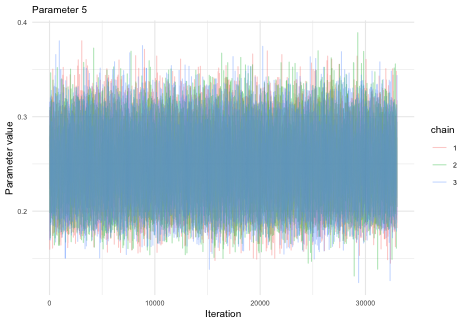} \includegraphics[width=0.24\linewidth]{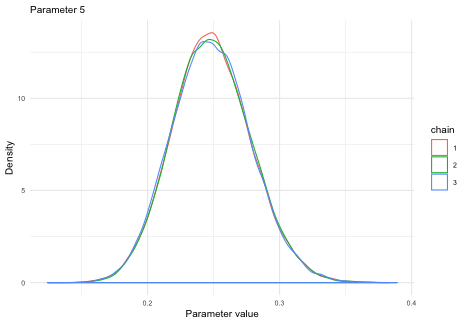} ~~\includegraphics[width=0.24\linewidth]{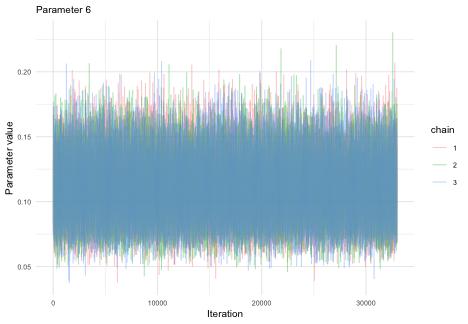} \includegraphics[width=0.24\linewidth]{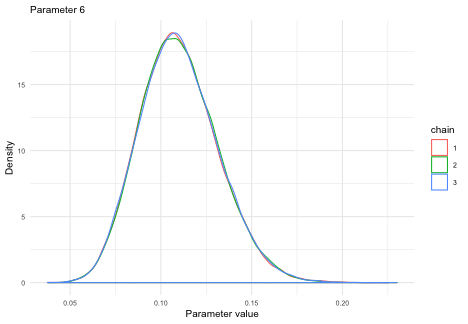} 
}

{\centering \includegraphics[width=0.24\linewidth]{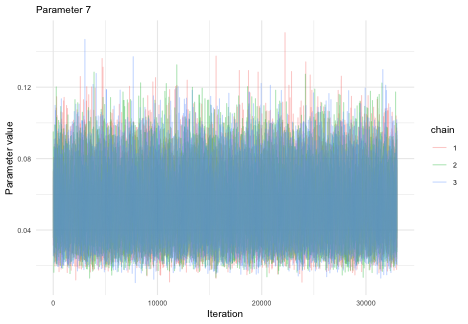} \includegraphics[width=0.24\linewidth]{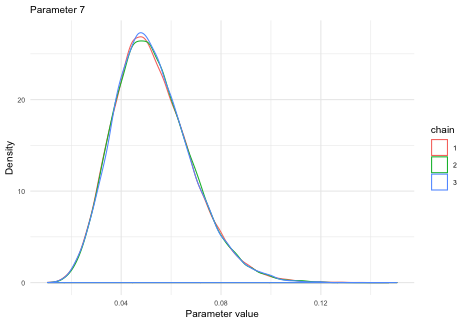} ~~\includegraphics[width=0.24\linewidth]{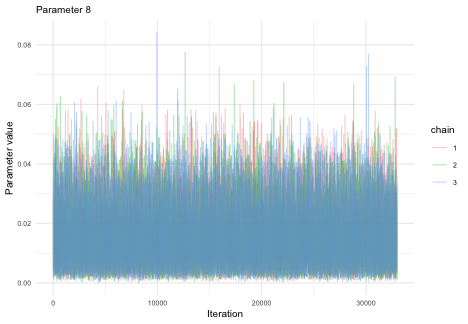} \includegraphics[width=0.24\linewidth]{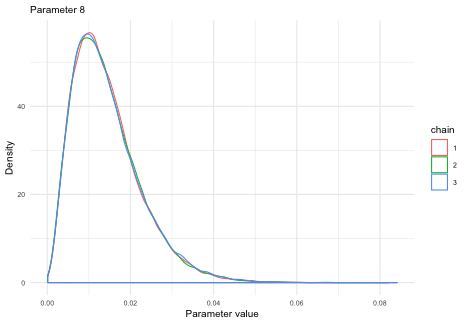} 
}

{\centering \includegraphics[width=0.24\linewidth]{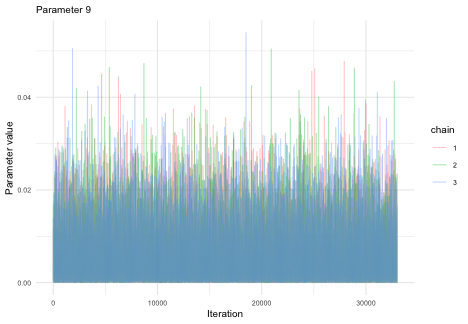} \includegraphics[width=0.24\linewidth]{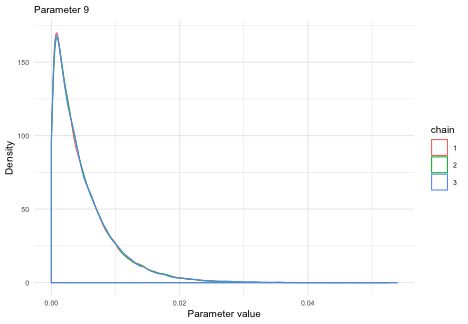} \includegraphics[width=0.24\linewidth]{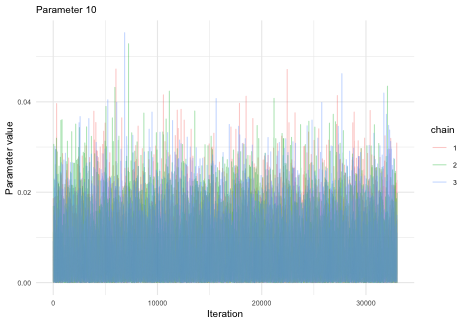} \includegraphics[width=0.24\linewidth]{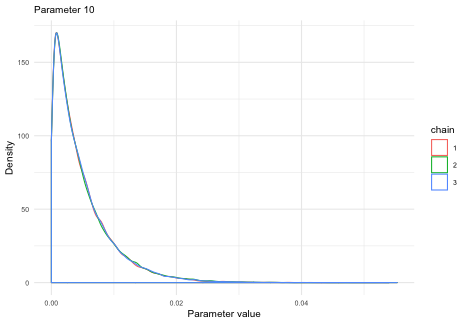}
}

\caption{Trace and density plots for each of the ten parameters and each of the three Markov chains we used to approximate the posterior distribution for \textbf{Final evaluation}. You can see that the MCMC simulations have reached convergence, and that states $9$ and $10$ are judged highly improbable as one would expect given the limited available data.}\label{fig:alphaPlots}
\end{figure}

In Figure \ref{fig:alphaPlots}, you can see the so called \emph{traceplot} and
\emph{densityplot} for each parameter of the categorical distribution of \textbf{Final evaluation}: \((p_1,\ldots, p_{10}) \sim \mathrm{Dir} (\alpha_1,\ldots, \alpha_{10})\), resulting from an MCMC simulation of our \texttt{chowliu} graph structure.

\begin{table}

\caption{\label{tab:nParams}Number of parameters associated to each node in the chosen Bayesian network (chowliu): the target feature has 10 possible states; other nodes have 5 states, but conditional dependences add dimensions to the CPTs.}
\centering
\begin{tabular}[t]{ll}
\toprule
node & \# of parameters\\
\midrule
FINAL\_EVAL & 10\\
PROP\_TO\_CHANGE & 5\\
LEAD\_INCL & 10x5\\
INTEGRITY & 10x5\\
VAL\_FOR\_CLI & 10x5\\
DEC\_MAKING & 10x5\\
INNOV\_SIMPL & 5x5\\
VISION & 5x5\\
COLLAB\_INFL & 25x5\\
\bottomrule
\end{tabular}
\end{table}

\hypertarget{conclusions}{%
\section{Conclusions}\label{conclusions}}

In this article, we presented a series of statistical tools that can be
effectively used in corporate context, in order to guide business decisions
and strategies. While each single element of the chain we used is well known 
in academic context, to our knowledge
they are seldom used together, to the full extent of their capabilities,
in industrial and strategic contexts. In presenting the specific experience
developed by the authors in the past few years, we hope to have conveyed the
wide range of possible applications of such methodology in different
scenarios.

In \protect\hyperlink{secEX}{Section \ref{secEX}}, we described a specific 
corporate situation where
the procedure leads to an effective support to domain experts in their job.
In the case at hand, the model allowed to improve experts' beliefs, which
were only partially confirmed about the importance of different features, as
well as to receive an independent score against which to ponder their
evaluation of human auditors.
This is a general trend of application for data science tools, and for
Bayesian models in particular: as human being, we are imperfect evaluator of
uncertain scenarios and even worse synthethizers of different quantitative
elements (especially due to heuristics and biases, see~\cite{ref-KahnemanTFS}). Hence,
having consistent and coherent ways to aggregate and synthetize different
sources of quantitative information, like the ones based on the Bayesian
approach to uncertainty, could be of paramount importance towards a more
effective course of actions in corporate decisions.

We would also like to stress the \emph{transparency} value of the Bayesian
techniques: in our context any manager could verify at any stage the effect
of the measured predictors (skills and attitudes of auditors) on the target
variable (global evaluation of auditors' job).
In different contexts, the same transparency allows managers to monitor
potential impacts of their strategies and to obtain a quantitative
assessment that they can use to explain and communicate reasons behind their
choices, e.g.~to stakeholders. This is made possible because 
Bayesian networks allow to analyze the whole predictive 
posterior distribution, through suitable marginalization over the unobserved features, offering
a more thorough measure of the impact of different features.

This is in stark contrast with most of the advanced machine learning
algorithms used in industry: AI tools based on \emph{deep neural networks} or
\emph{(extreme) gradient boosting} techniques, just to name two of the most
popular algorithms, are typically ``black boxes'' that only provide a single
prediction or probability, without any mean to unroll the sequence of
evaluations which led the trained system to its result. Despite the recent
efforts to develop \emph{local explainers}, like the \emph{Local Interpretable
Model-agnostic Explanations} (or \emph{LIME}) approach described in~\cite{ref-LIME}, or 
local and global \emph{SHAP values} described in~\cite{ref-SHAP}, the
search for the reasons behind predictions given by black box algorithms is
still way more difficult than in the case of Bayesian tools and it offers
only partial results. The price to pay is in higher computational requirements than 
what would be needed for approximated measures like the aforementioned 
SHAP values~\cite{ref-SHAP}, or partial dependence plots~\cite{ref-BBOXzh}.
Nonetheless, we believe that the insights gained through the inspection of
complete distributions is usually worth the effort and, thus, that the Bayesian 
approach can really represent an extremely powerful support towards more 
rational \emph{data-driven decisions} in any corporate context.

\vspace*{12pt}

\noindent \textbf{Acknowledgement:} The authors thank Prof. Julia Mortera for the fruitful discussions during the preparation of the paper. They also thank the anonymous referee
for various suggestions which helped to improve the final revision of the paper.

\hypertarget{appTech}{%
\section{Appendix - Technical aspects}\label{appTech}}

In this section we focus on some technical ingredients which we have built our tools upon. They represent a selection of the rich catalogue of techniques for the so-called \textbf{Bayesian inference and data analysis}~\cite{ref-JaynesBook, ref-GelmanBDA, ref-KrushkeBDA}: Bayesian networks, entropy and mutual information among variables, structure learning, Bayes-Turing factor and MCMC simulations.

Bayesian techniques are invaluable in corporate processes, because they allow to exploit all available information, usually quite sparse, and to synthetize it consistently with experts' opinions that can be modelled, not without considerable efforts, in forms of prior distribution of probability to be inserted into the model.

Since we are not adding new statistical developments to such tools and techniques, however, we recall here only some of them, for completeness sake.
Readers, interested into general Bayesian networks and other topics briefly touched in this section, shall look into the aforementioned books and in~\cite{ref-CowellPNES, ref-PearlCausality, ref-HELgmodel}.

\hypertarget{bayesian-networks}{%
\subsection{Bayesian networks}\label{bayesian-networks}}

A Bayesian network is a \emph{directed acyclic graph} (DAG) whose nodes represent measurable features, and whose links between nodes represent mutual influences among variables. Since links are directed, such a structure creates a hierarchy among nodes and we will call \textbf{parent nodes} the starting points of such links and \textbf{child nodes} their ending points.

In order for any inference to take place, each variable has to be \emph{modelled} in some way, i.e.~a probability distribution of the variable has to be assumed, and each directed link is to be interpreted as a conditional dependence of the \emph{child} variable on the \emph{parent} one.

When dealing with discrete (or discretized) features only, as in the specific
applications presented in \protect\hyperlink{secEX}{Section \ref{secEX}}, nodes of the graph are
endowed with categorical distributions, and their parameters are assumed to
be distributed according to suitable Dirichlet distributions. By combining
structure learning algorithms (which will be presented afterwards) with
domain experts' knowledge, it is then possible to figure out the most
appropriate network structure and to impose prior distributions to the
parameters.
When models involve continuous features, like price or reliability estimates, the effort
is higher because one needs some rigorous justification to impose the specific
distributions.
However, in the long run, we believe the resulting predictions definitely pay back the
initial efforts, thanks to their explainability and to the production of a complete
\emph{a posteriori} distribution for each parameter. This, in turn, allows more thorough
analysis of the phenomena, compared to what is possible with standard point estimates.

\hypertarget{variable-selection-through-mutual-information}{%
\subsection{Variable selection through mutual information}\label{variable-selection-through-mutual-information}}

The very first step to be performed, when building a model of the corporate
process of interest, is the analysis of which features, among the available
ones, are truly relevant in monitoring and controlling the process.
In this section, we assume that available measures already monitor all
aspects of interest and we focus our attention to the actual selection of
relevant variables vs.~redundant ones.

In our procedure, we choose to select relevant features by measuring
the reciprocal influences between pairs of predictors, in a way that is
independent of the model structure. As a metric to quantify such influences,
we introduce the concept of \emph{mutual information}, taken from information theory, and use such metric to decide which variables shall be kept and which discarded. In order to define
such a quantity, we first borrow the concept of \emph{(Shannon) entropy} : given a random variable \(X\) taking discrete values
in the set \(\mathcal{X}\), we define its entropy as
\begin{equation}\label{eq:entropy}
H(X) = -\sum_{x\in \mathcal{X}} ~p(x)~ \log p(x)~.
\end{equation}
The idea beyond this quantity is to measure the amount of information about the
distribution of \(X\) that we can gather from the observation of evidences, given how probable
each observation is: rarer events will provide more information; more common ones will
provide much less information. A conditional version \(H(X|Z)\) of the entropy can be defined analogously, for a pair of random variables \(X\) and \(Z\), by simply using the conditional distribution \(p(x|z)\) in place of \(p(x)\).

Once entropy has been defined, we can also give the following definition.
If \(X,Y\) are random variables supported on discrete sets \(\mathcal{X}\) and \(\mathcal{Y}\), we define \textbf{mutual information} \(\mathrm{MI}\) among \(X\) and \(Y\) the quantity
\begin{equation}\label{eq:mutinfo}
\mathrm{MI}(X,Y) = H(X) + H(Y) - H(X,Y)~,
\end{equation}
where \(H(X,Y)\) is the entropy of the joint distribution of \((X,Y)\). Such a measure offers a few advantages compared to \emph{(Pearson's) correlation coefficient} and its extensions: it gives a more appropriate measure of the relation between categorical variables \(X\) and \(Y\), whenever one of the features does not map well in terms of a numerical representations; it can be extended in a natural way to cope with the case of conditional probability distributions. This latter aspect plays a crucial role in selecting pairs of variables that can be combined to produce a Bayesian model of the corporate process.

Therefore, we will also define, for any triple of categorical variables \(X,Y,Z\) taking values, respectively, in \(\mathcal{X},\mathcal{Y}, \mathcal{Z}\), the \textbf{conditional mutual information} \(\mathrm{CMI}\) as follows
\begin{equation}\label{eq:condmi}
\mathrm{CMI}(X,Y|Z) = 
H(X|Z) + H(Y|Z) - H(X,Y|Z)~.
\end{equation}
It is possible to give the same definition in presence of multiple conditioning on variables \(Z_1,\ldots,Z_k\), by simply replacing the distributions conditional on \(z\in \mathcal{Z}\) with those conditional on multiple variables.

However, there is still a detail to be accounted for: different pairs or triples of random variables cannot always be directly compared through their \(\mathrm{MI}\) and \(\mathrm{CMI}\).
When we want to compare different pairs of variables \(X,Y\) and \(Z,W\), we might end up with
\[
\mathrm{MI}(X,Y) \ll \mathrm{MI}(Z,W)
\]
simply because variables \(Z\) and \(W\) have a different distribution compared to \(X\) and
\(Y\), that gives them larger entropies, despite the influence of \(Y\) on \(X\) being
possibly very important in the model of the process.
This situation is quite common in corporate applications. For instance, you can think to a binary classification problems with rare class of interest (like in fraud or churn management processes): if other variables are distributed in a more uniform way, it might well be that scores involving the target variable can be order of magnitude smaller than scores not involving it.
In order to cope with these cases, without the need of discriminating among variables
with high and low entropy, we decided to replace
\(\mathrm{MI}(X,Y)\) from \eqref{eq:mutinfo} and \(\mathrm{CMI}(X,Y|Z)\) from \eqref{eq:condmi} with their normalized versions~\footnote{Sometimes
  \(\mathrm{MI}'\) is called \emph{symmetric uncertainty}, or \(\mathrm{SU}\). By analogy, we could
  call \(\mathrm{CMI}'\) \emph{conditional symmetric uncertainty}.}
\begin{equation}\label{eq:normalized}
\mathrm{MI}'(X,Y)~ =~ 2~ \frac{\mathrm{MI}(X,Y)}{H(X)+H(Y)}\,,
\qquad\qquad
\mathrm{CMI}'(X,Y|Z)~ =~ 2~ \frac{\mathrm{CMI}(X,Y|Z)}{H(X|Z)+H(Y|Z)}\,.
\end{equation}

\begin{figure}

{\centering \includegraphics[width=0.33\linewidth]{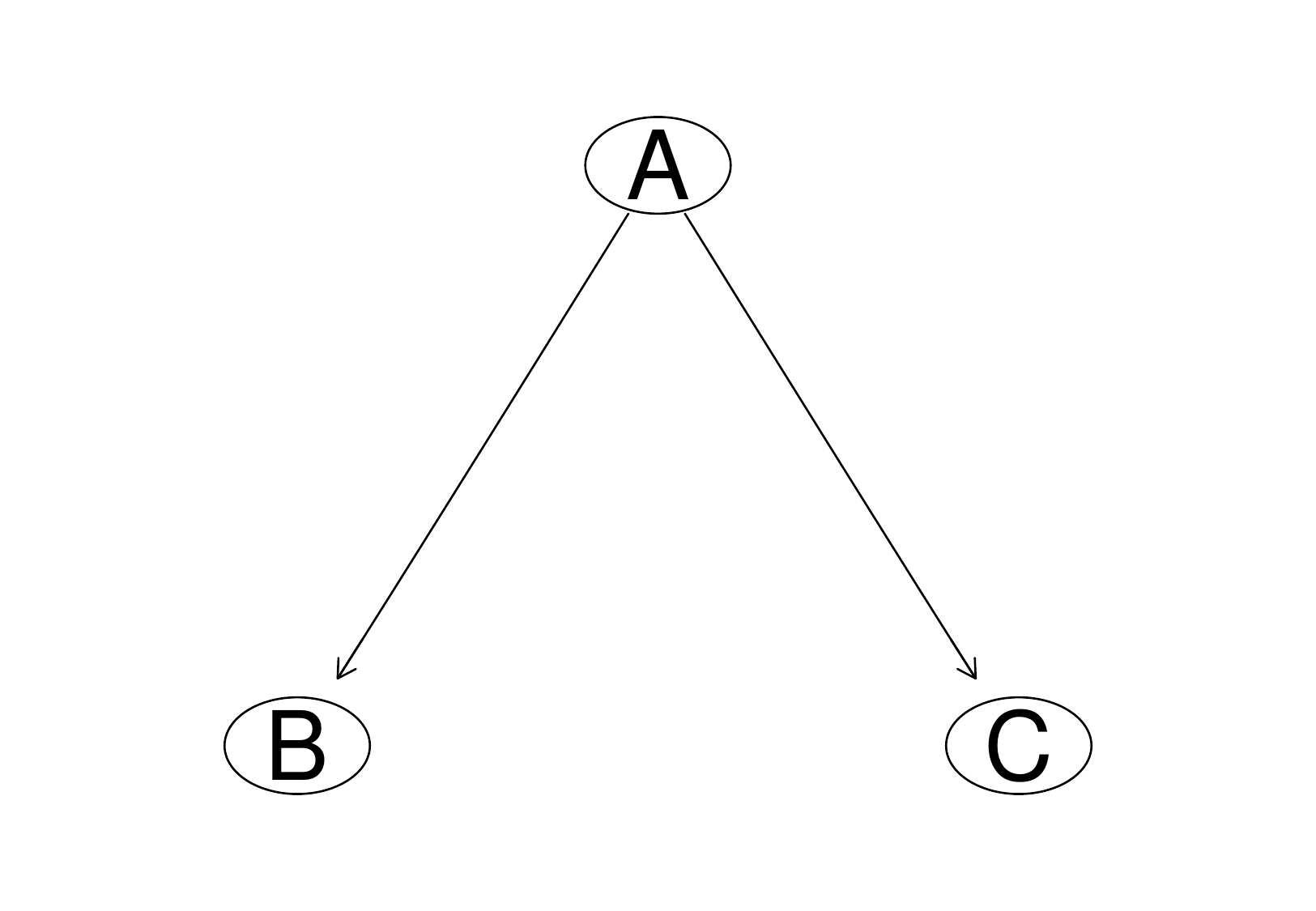} 

}

\caption{Features $B$ and $C$ could have effect on each other only when conditioned on their common parent $A$: in view of the construction of a Bayesian network, it is important to explore conditional dependencies.}\label{fig:Vrevfig}
\end{figure}

With these measures at hand, the variable selection procedure can be carried out through a careful analysis of the quantities \(\mathrm{MI}'(X,Y)\) and \(\mathrm{CMI}'(X,Y|Z)\), as \(X,Y,Z\) vary over all the available features.
When variables have high \(\mathrm{MI}'\), they shall be considered as candidate connections in a Bayesian model. Also, pairs which have high \(\mathrm{CMI}'\) when conditional on a third variable, shall be evaluated in a Bayesian network as forming a graph structure as the one depicted in Figure \ref{fig:Vrevfig}. Finally, there might be value in the addition to the Bayesian model of triplets of variables \(X,Y,Z\) which have \(\mathrm{CMI}'(X,Y|Z)\gg \mathrm{MI}'(X,Y)\), especially when either of the variables \(X,Y\) is of particular interest according to the experts of the corporate process.

Notice that similar analysis are exploited by structural learning algorithms for Bayesian networks (see e.g.~\cite{ref-ScutariBook, ref-MargaritisPHD}) and they are used to analyse possible causal relationship in a polytree or a DAG (see~\cite{ref-PearlCausality}).

\hypertarget{structure-learning-algorithms-for-bayesian-networks}{%
\subsection{Structure learning algorithms for Bayesian networks}\label{structure-learning-algorithms-for-bayesian-networks}}

Once we have selected candidate features for inclusion in our Bayesian
model, our next step focuses on inferring the actual structure of the model.
This means to infer, from the available data, a DAG structure whose
connections represent in the best possible way the relationship among
predictors and targets.
In order to perform such a step, we have to move beyond the local evaluation
of dependence among pairs (or triplets) of variables and to compute the
likelihood of our dataset of evidences, subject to the different DAGs which
can connect the variables.
Hence, in this section we move our attention towards algorithms developed to perform \emph{structure inference}, such as the ones described in~\cite{ref-MargaritisPHD} and implemented in some R packages like \texttt{bnlearn}~\cite{ref-ScutariBN, ref-ScutariBook}, \texttt{bnclassify}~\cite{ref-BNCL}, and \texttt{BNSL}~\cite{ref-SuzukiBN}.

Despite a certain variety of theoretical motivations, most if not all such algorithms can be reduced to three broad categories: \textbf{score-based} algorithms, \textbf{constraint-based} algorithms, and \textbf{hybrid} algorithms.

The first category, \emph{score-based} algorithms, approaches the problem by searching a DAG structure which maximizes a certain score measuring the posterior probability of each DAG structure \(\mathcal{G}\) given the dataset \(\mathcal{D}\).
Since
\[
P(\mathcal{G}|\mathcal{D}) ~\propto~ P(\mathcal{D}|\mathcal{G})~P(\mathcal{G})
\]
most algorithms assume that \(P(\mathcal{G})\) is uniform over a set of possible structures and concentrate on finding an approximate maximum of \(P(\mathcal{D}|\mathcal{G})\).
However, the whole space of admissible DAGs is too large to be fully
explored and, typically, \emph{greedy search} algorithms are used: one starts from a given initial network structure and it tries to modify it (by adding, deleting or reversing one arc at time), until there is no further change that improves the score. We will see the action of a score-based algorithm through the \texttt{hc} function of \texttt{bnlearn}.

\begin{figure}
\includegraphics[width=0.3\linewidth]{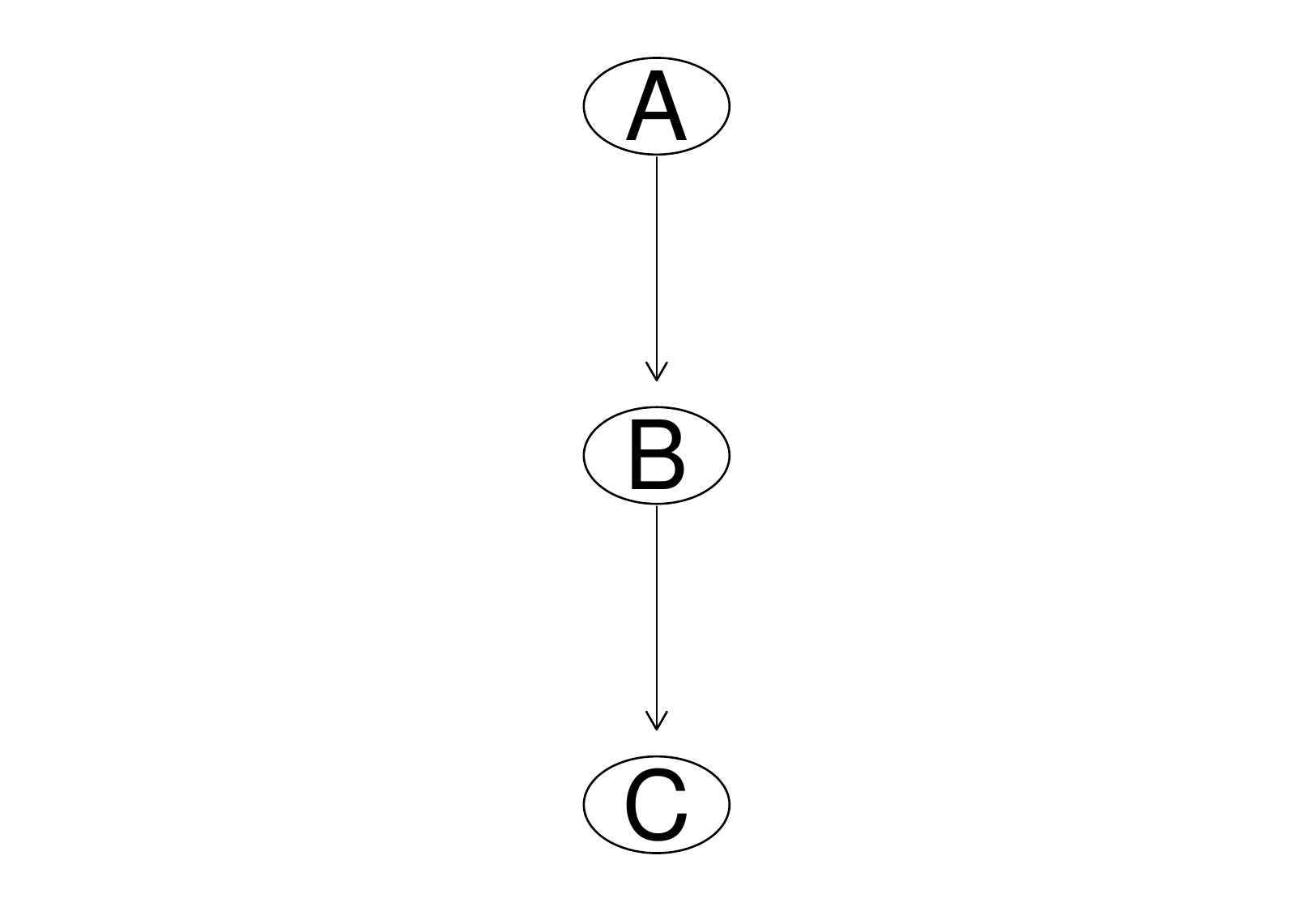} \includegraphics[width=0.3\linewidth]{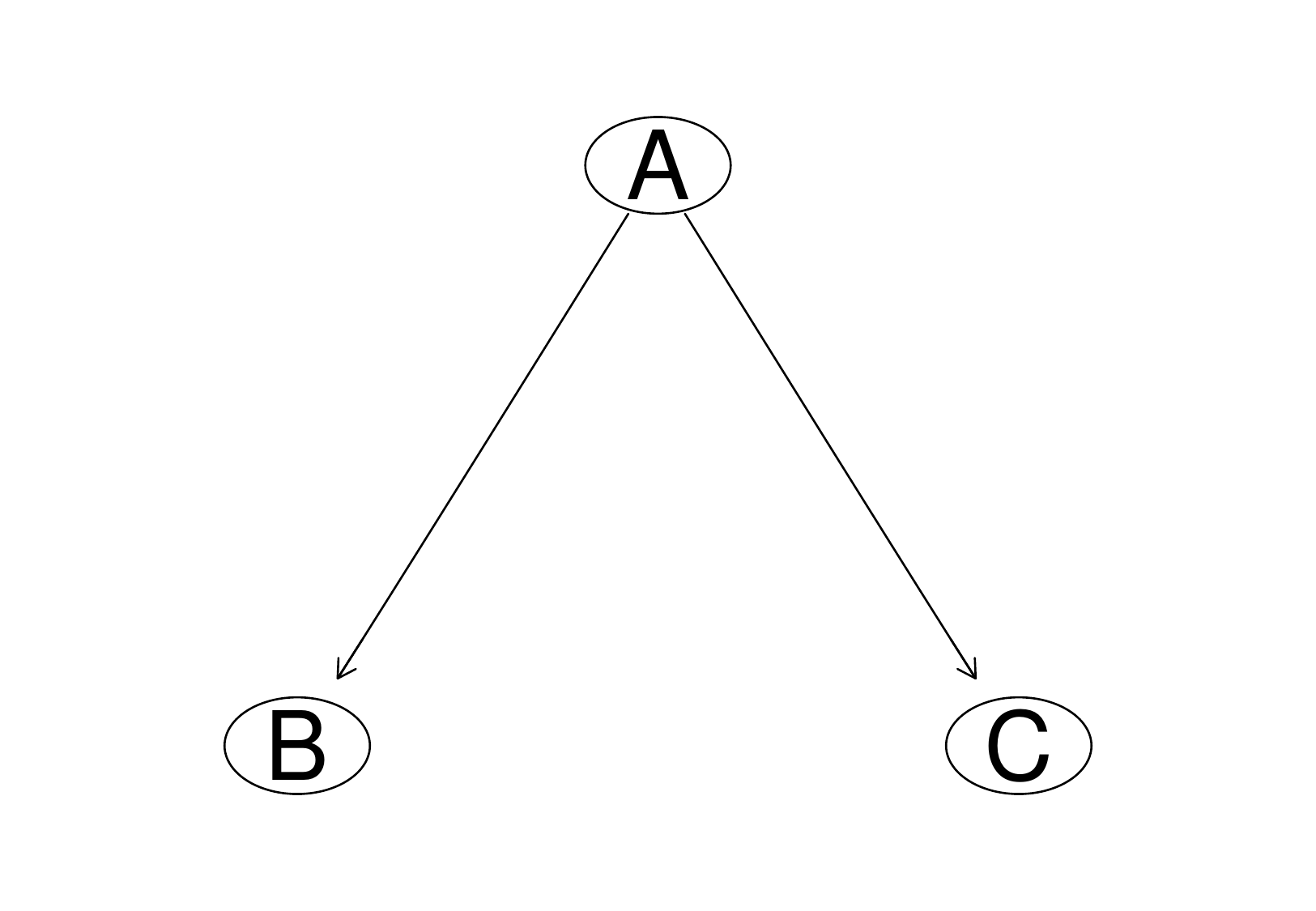} \includegraphics[width=0.3\linewidth]{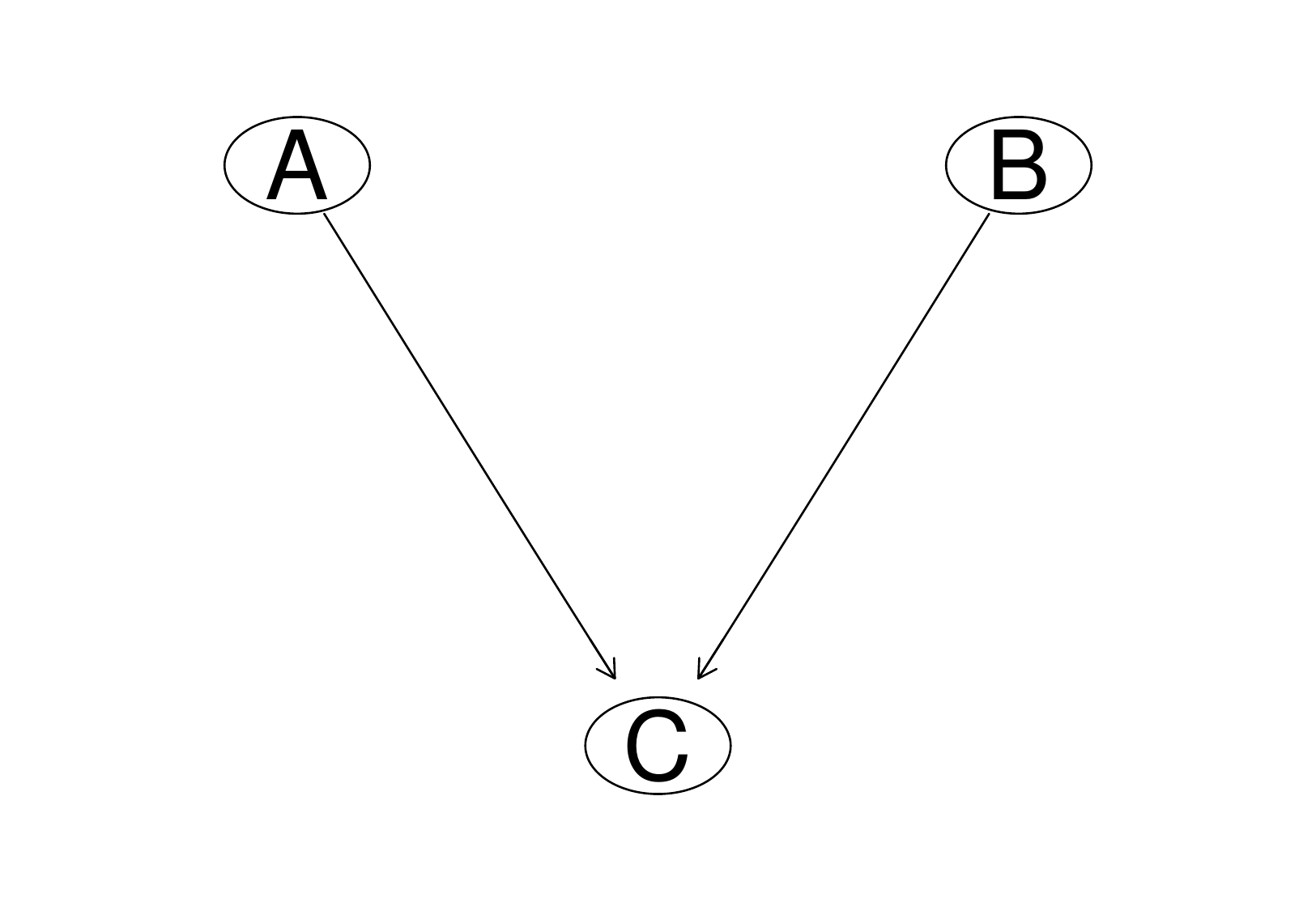} \caption{Different types of causal sub-structures that can be assigned by constraint-based structure learning algorithms.}\label{fig:causeSubstr}
\end{figure}

\emph{Constraint-based} structure learning algorithms, on the other hand, proceed along different lines. They were initiated by the work by Verma and Pearl~\cite{ref-VermaPearl} and attempt to:

\begin{itemize}
\tightlist
\item
  first identify which pairs of variables shall be connected by an arc, regardless of its direction,
\item
  and then assign directions to arcs by identifying the correct \(V\)-structures among all nonadjacent pair of nodes with a common neighbor (cfr. Figure \ref{fig:causeSubstr}).
\end{itemize}

More precisely, in the former phase, algorithms of this kind test for the existence of
connections by means of some independence test or information theory score (e.g.~
\(\mathrm{MI}\) or \(\mathrm{MI}'\) as defined in \eqref{eq:mutinfo} and \eqref{eq:normalized}, cfr.~\cite{ref-Cheng97}),
while in the latter phase they exploit conditional versions of the same test or score
to search for the appropriate structures. We will see the action of a constraint-based algorithm through the \texttt{chow.liu} function of \texttt{bnlearn}.

Finally, \emph{hybrid} structure learning algorithms have been proposed by combining aspects from constraint-based algorithms and aspect score-based ones. They typically proceed in two steps, called \emph{restrict} and \emph{maximize}: they start by finding a candidate set \(\mathcal{P}_i\) of parents for each node \(N_i\) (selected among nodes that are somehow `related' to \(N_i\)), then they search the graph \(\mathcal{G}\) which maximizes a given score, among structures that only contain edges connecting \(N_i\) with nodes in \(\mathcal{P}_i\).

\hypertarget{model-selection-through-bayes-turing-factor}{%
\subsection{Model selection through Bayes-Turing factor}\label{model-selection-through-bayes-turing-factor}}

After applying the \emph{structure learning} algorithms described in the previous section, we end up with many possible candidate models to describe the data-generating corporate process. Since our final goal consists of selecting a
single model, which has captured most salient characteristics of the process and thus is capable to correctly predict
future values of the \emph{target variables}, we need a way to evaluate how well each model matches the available
training data.

The approach we followed is the so-called \emph{Bayesian model selection}, which builds a rank of the different models based on the \textbf{Bayes-Turing factor} (denoted by \(\mathrm{BF}\) later on). Let our competing models be denoted by \(\mathcal{H}_1\) and \(\mathcal{H}_2\), and let \(\mathcal{E}\) be the given set of evidences. Then, we define the \emph{Bayes-Turing factor} as
\begin{equation}\label{eq:BF}
\mathrm{BF}(\mathcal{H}_1, \mathcal{H}_2) = \frac{P(\mathcal{E}|\mathcal{H}_1)}{P(\mathcal{E}|\mathcal{H}_2)}\,.
\end{equation}

It is easy to see that, since \(P(\mathcal{H}_i|\mathcal{E}) \propto P(\mathcal{E}|\mathcal{H}_i)\,P(\mathcal{H}_i)\), \(\mathrm{BF}(\mathcal{H}_1, \mathcal{H}_2)\) is exactly the factor that allows to update the ratio
\(P(\mathcal{H}_1)/P(\mathcal{H}_2)\) when new
evidences are collected.

In our context, we exploit the following facts:

\begin{itemize}
\tightlist
\item
  a value \(\mathrm{BF}(\mathcal{H}_1, \mathcal{H}_2)>1\) means that the evidences increase the chances of model \(\mathcal{H}_1\) vs.~model \(\mathcal{H}_2\), compared to prior knowledge,
\item
  a value \(\mathrm{BF}(\mathcal{H}_1, \mathcal{H}_2)<1\) means that the evidences favors model \(\mathcal{H}_2\) over model \(\mathcal{H}_1\), compared to prior knowledge,
\item
  a value \(\mathrm{BF}(\mathcal{H}_1, \mathcal{H}_2)=1\) means that the evidences do not help discriminating among the models in any way.
\end{itemize}

\begin{figure}

{\centering \includegraphics[width=0.33\linewidth]{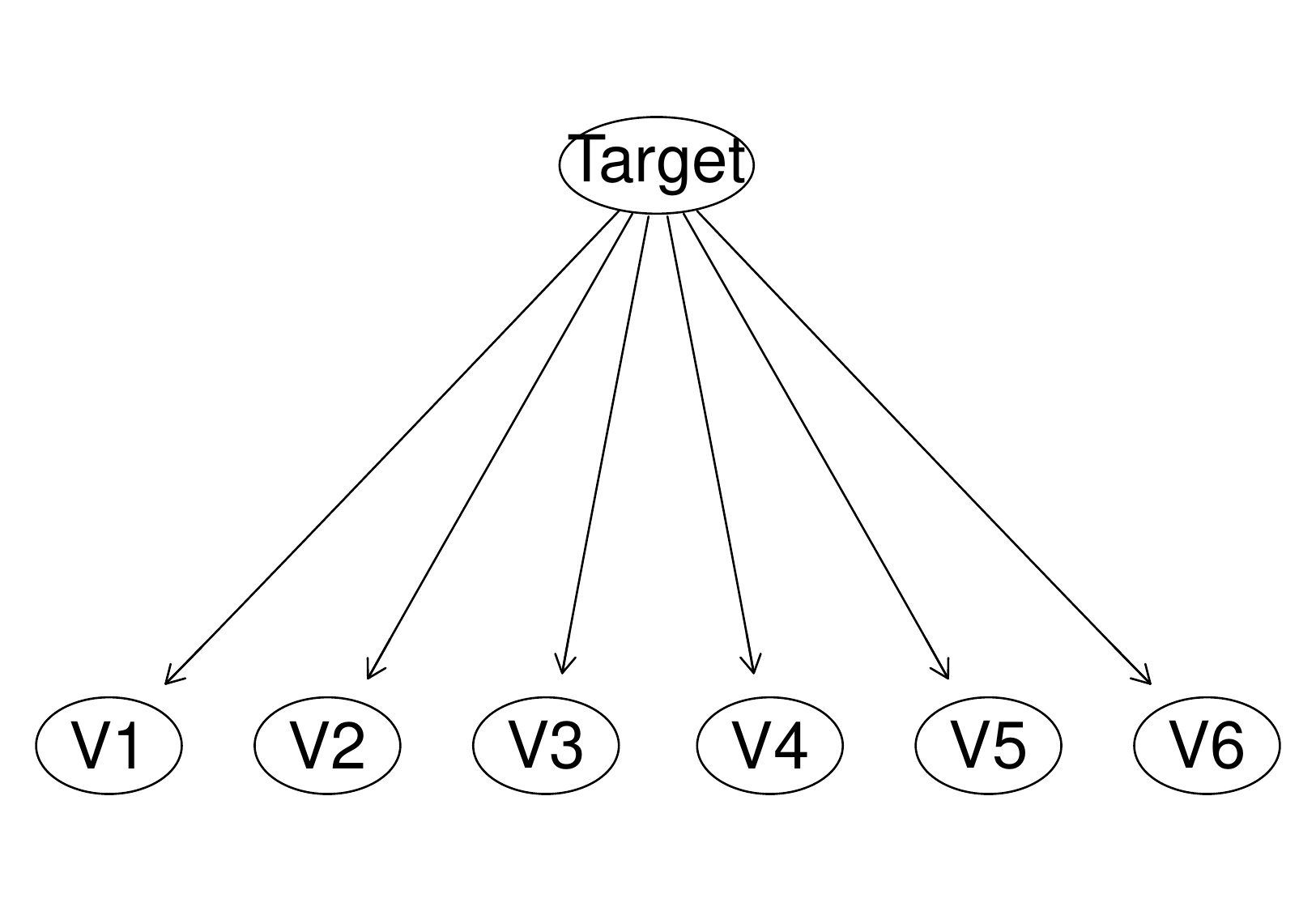} 

}

\caption{Na\"{\i}ve structure of Bayesian network: the target variable is the only parent node and all regressors are conditioned to it, while being (conditionally) independent from each other.}\label{fig:NBfig}
\end{figure}

By adopting BF in our workflow, we can effectively rank the different graph models suggested by \emph{structure learning} algorithms and select the most promising one as the model to be used to support the decision making processes.
Moreover, it becomes very easy to also add a \emph{na\"{\i}ve Bayes} graph (cfr. Figure
\ref{fig:NBfig}) in the ranking, i.e.~a network where the target variables are parent
nodes of all remaining features and no links are present among predictors. This simple structure acts as a reasonable benchmark to quantify how much the additional complexity, given by a non-na\"{\i}ve structure,
actually pays off compared to the cheaper na\"{\i}ve model.

\hypertarget{inference-of-parameters-through-mcmc-simulations}{%
\subsection{Inference of parameters through MCMC simulations}\label{inference-of-parameters-through-mcmc-simulations}}

We got to the point where we have exploited our dataset to automatically select the
variables with a major impact on the target variable, and to design the most likely DAG
model for the data-generating process.
The result of this sequence of operations is a model of the corporate
process, depending on suitable parameters coming from the probability
distributions associated to each node of the DAG. Our goal, at this stage,
is to perform a Bayesian inference of all such parameters, so to obtain
an estimate which incorporates both our prior beliefs and the
evidences collected through data.

However, in general, the distributions in the model are not conjugate pairs of prior and
likelihood, for which we could infer the parameters in closed analytical forms. This
means that the posterior distributions has to be simulated via \emph{MCMC} methods. Such an approach, originated in 1950s with
Metropolis-Hastings algorithm~\cite{ref-mhMC} with the goal
of approximating distributions of particle collisions in Physics, allows to construct a
\emph{sampler} which produces collections of simulated data, whose distribution approximates
accurately (on the long run) the underlying probability distribution.

\begin{figure}

{\centering \includegraphics[width=0.3\linewidth]{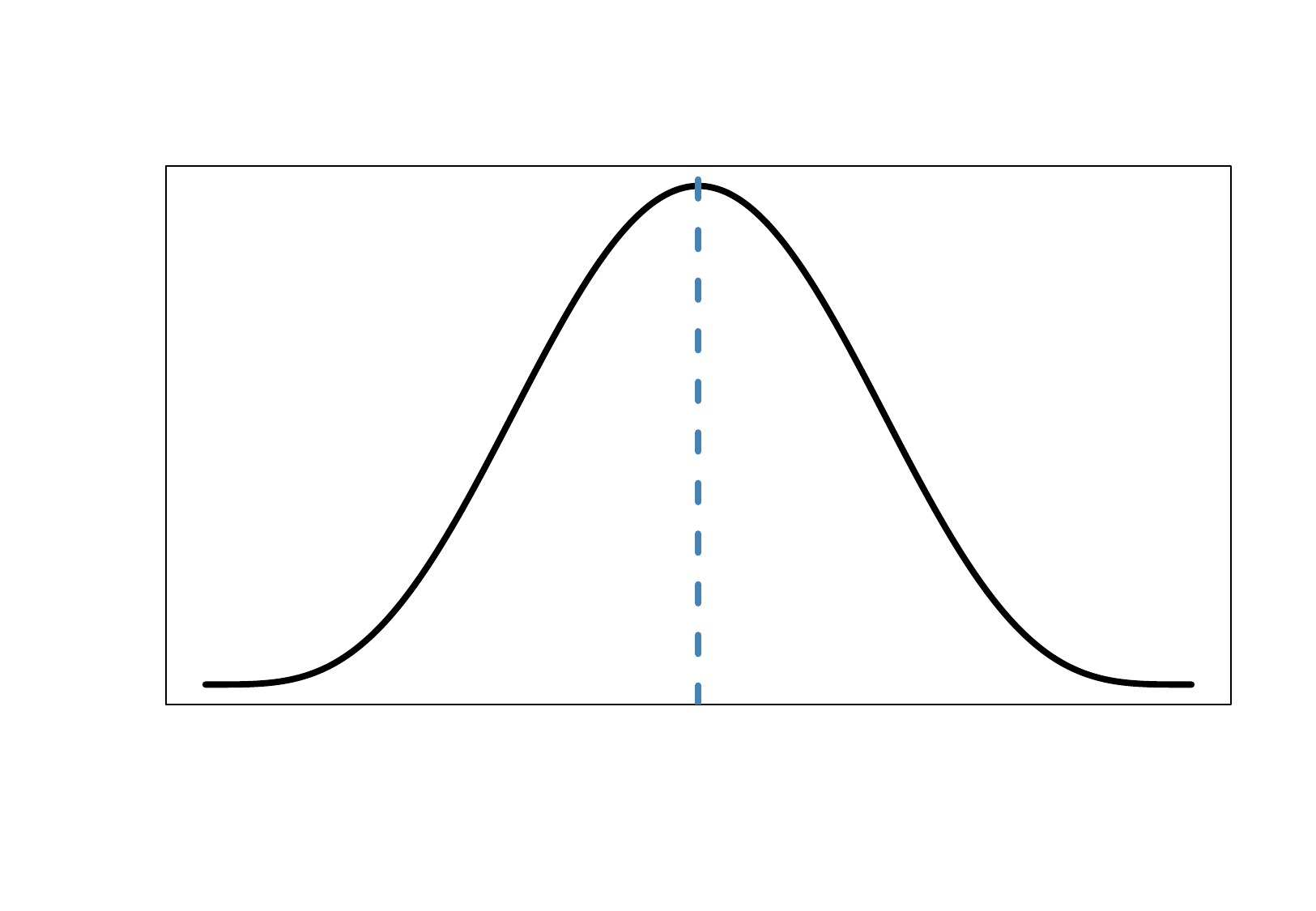} \includegraphics[width=0.3\linewidth]{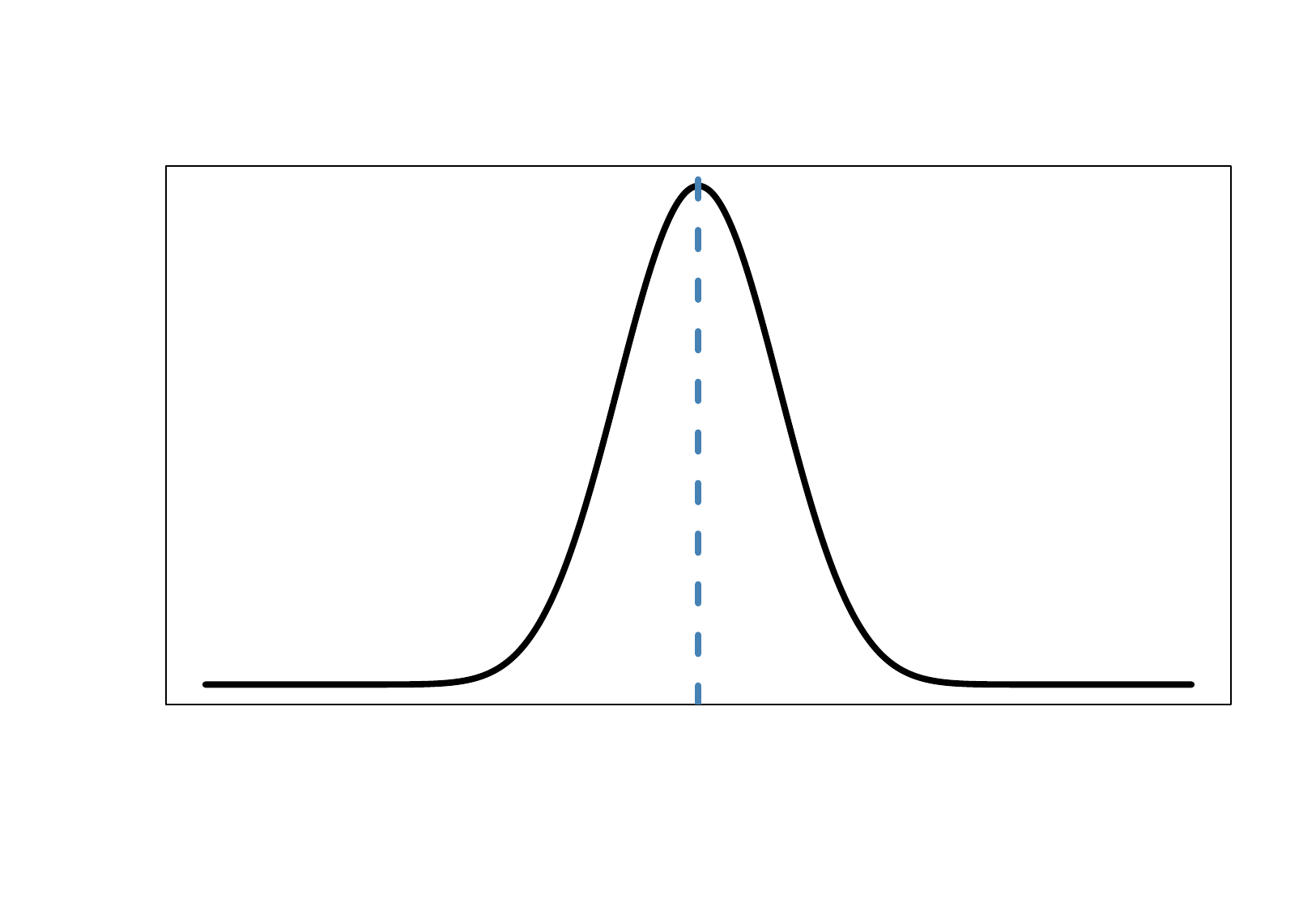} \includegraphics[width=0.3\linewidth]{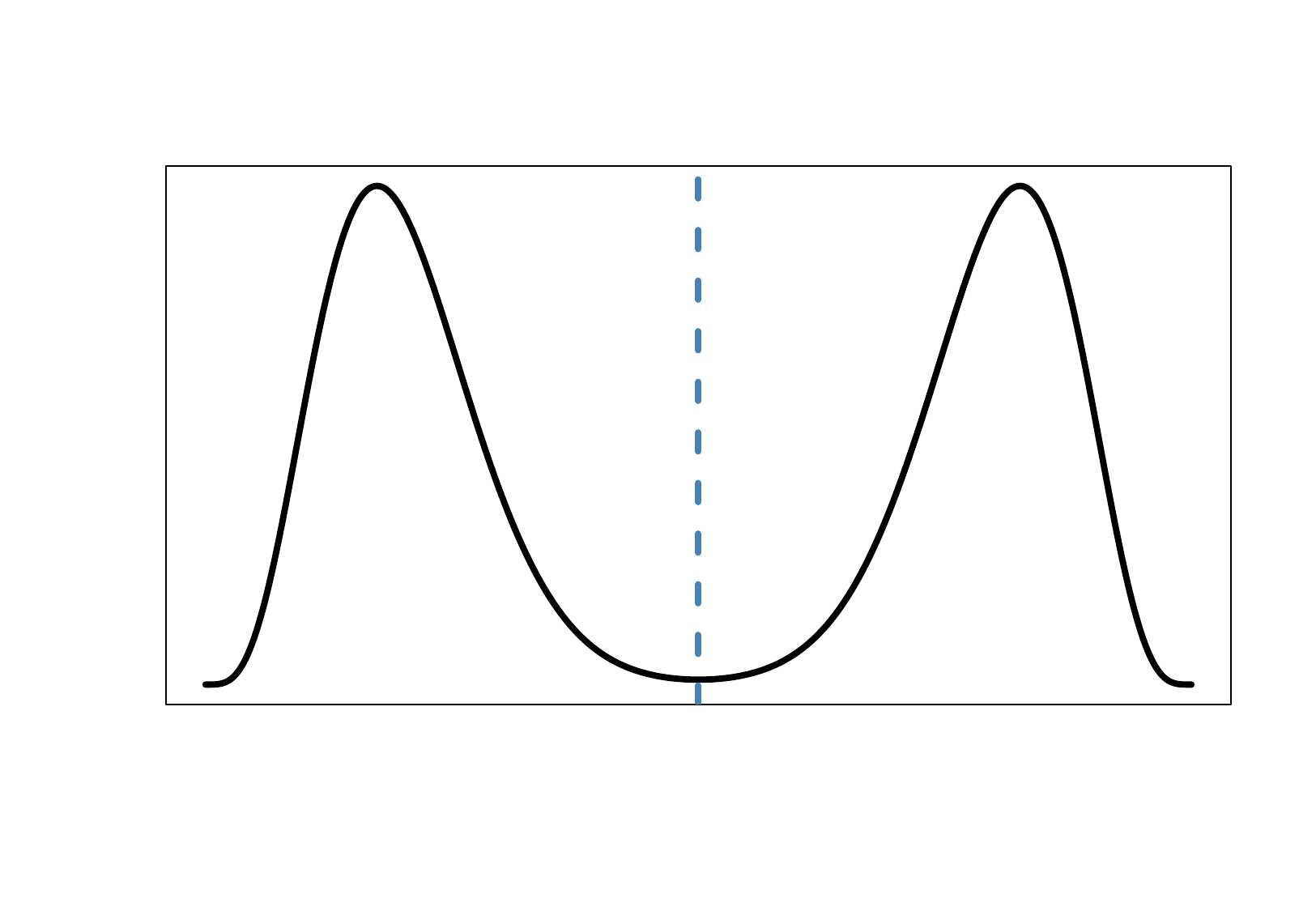} 

}

\caption{Some probability densities have differences that could not be summarized by a single pointwise estimate, e.g.~the location of the mean of the posterior (dashed line).}\label{fig:mixDistr}
\end{figure}

In our corporate use case, this set of techniques allows to approximate very
accurately the posterior distribution of the parameters, instead of
obtaining only a pointwise estimator, like we could get with traditional
\emph{maximal likelihood estimates} (\emph{MLE}) or with a
\emph{maximum a posteriori} (\emph{MAP}) estimate. In turn, this enables experts from
the corporate domain to judge by themselves how good a pointwise
approximation would be, e.g.~by discriminating
among the distributions in Figure \ref{fig:mixDistr}.

We leave additional technical details to specific references in the Bibliography~\cite{ref-GelmanBDA, ref-KrushkeBDA, ref-gsTUT}.

\hypertarget{sequential-usage-of-the-different-components}{%
\subsection{Sequential usage of the different components}\label{sequential-usage-of-the-different-components}}

Now that we have outlined separately each step of the methodological approach, it is time to summarize them all in a more comprehensive way.

Let us consider an hypothetical expert in a corporate environment, which is asked to decide the course of action to improve some prescribed KPI, based on a large dataset where several historical indicators have been recorded.
According to the procedure streamlined in the previous sections, our expert will first perform a selection of the most significant predictors, based on \(\mathrm{MI}'\) and \(\mathrm{CMI}'\). Here, she will have a first chance to exploit her expertise into the process: specific features, deemed of interest, can be ``manually'' incorporated even if they obtained a low score, based on available data. This allows to include in later analysis both the experience of human decision makers \textbf{and} the \emph{signal} present in the data, that could be driven by newer phenomena or by unexpected features of the process.

Then, the expert would feed the selected features into the component which performs
structure learning, so to obtain different proposals of Bayesian models trained over
the dataset. Here, she can prescribe the inclusion or exclusion of specific connections in the construction performed by algorithms, based on her judgement.
In the end, by comparing of the Bayes-Turing factor \eqref{eq:BF} among pairs of different
structures, she can select one or a few of them that seem valuable and that will be ultimately compared and ranked on the base of their actual predictive capabilities.

Finally, the chosen model can be trained and put in action, through MCMC simulations, so to predict future values of the target variable.

Notice that, in order to mitigate the risk of \emph{overfitting} the training data
with the structure inference and the parameter inference, one might want to
perform the final selection through a \(k\)-fold cross validation procedure~\cite{ref-BayesCV}.
This is particularly useful in the case where multiple network structures
offer similar performances at the end of the structure learning phase, 
i.e.~they have a Bayes factor close to 1: the numerical simulations can
then be used to evaluate more in details the generalization capabilities of
each model, by selecting the structure which offers the best average
performances among the different validation sets. The performance metric
shall be chosen based on the specific problem at hand, either as a measure
of the prediction errors or as a quantification of costs associated to
predictions.
More complex evaluation procedures could be set up as well (e.g.~iterated
cross validation, nested cross validation, etc.), when desired, but
we do not account for them in this paper.

\end{document}